\documentclass[aps,floats,amssymb,twocolumn]{revtex4}
\usepackage{calc}
\usepackage{psfrag}
\usepackage{graphicx}
\usepackage{color}

\begin{document}

\begin{abstract}
We study the static structure factor of the fractional Chern insulator Laughlin-like state and provide analytical forms for this quantity in the long-distance limit. In the course of this we identify averaged over Brillouin zone  Fubini Study metric as the relevant metric in the long-distance limit. We discuss under which conditions the static structure factor will assume the usual behavior of Laughlin-like fractional quantum Hall system i.e. the  scenario of Girvin, MacDonald, and Platzman [Phys. Rev. B 33, 2481 (1986)]. We study the influence of the departure of the averaged over Brillouin zone Fubini Study metric from its fractional quantum Hall value which appears in the long-distance analysis as an effective change of the filling factor. According to our exact diagonalization results on the Haldane model and  analytical considerations we find persistence of fractional Chern insulator state even in this region of the parameter space.
\end{abstract}
\title{On the geometrical description of fractional Chern insulators based on static structure factor calculations}
\author{E. Dobard{\v{z}}i\'c$^1$}
\author{M.V. Milovanovi\'c$^2$}
\author{N. Regnault$^{3,4}$}
\affiliation{$^1$ Faculty of Physics, University of Belgrade, 11001 Belgrade, Serbia\\
$^2$ Scientific Computing Laboratory, Institute of Physics
Belgrade, University of Belgrade, Pregrevica 118, 11 080 Belgrade,
Serbia\\
$^3$ Department of Physics, Princeton University, Princeton, NJ 08544\\
$^4$ Laboratoire Pierre Aigrain, ENS and CNRS, 24 rue Lhomond, 75005 Paris, France}

\maketitle

\section{Introduction}\label{Section:Introduction}
Chern insulators (CIs)~\cite{haldane-1988PhRvL..61.2015H} exhibit integer quantum Hall effect (IQHE) conductance quantization in the absence of the magnetic field  due to non-trivial filled band structure with non-zero topological Chern number. Fractional Chern insulator (FCI)~\cite{neupert-PhysRevLett.106.236804,sheng-natcommun.2.389,regnault-PhysRevX.1.021014} is the name for CI with a partially filled band (which is akin to a Landau level) in the presence of strong interactions, which exhibit fractional quantum Hall effect (FQHE) conductance quantization.

What might be defined as an ultimate goal in the context of the FCI physics would be the understanding of the mechanism of creation of FCI states in order to be able to suggest the most convenient experimental settings, whether in cold atoms, crystal (solid-state) physics, or graphene structures for their realization.  Interacting topological insulator~\cite{Hasan-RevModPhys.82.3045,Qi-RevModPhys.83.1057,Qi-2010PhT....63a..33Q} physics, which FCI is a time-reversal-symmetry-broken representative, is one of the major topics of the current research in the field of strongly correlated systems. FCIs have been reported in many models both for fermionic systems~\cite{neupert-PhysRevLett.106.236804,sheng-natcommun.2.389,regnault-PhysRevX.1.021014,Venderbos-PhysRevLett.108.126405} and bosonic cold atom models~\cite{wang-PhysRevLett.107.146803,Yao-PhysRevLett.110.185302,Hafezi-PhysRevA.76.023613,Moller-PhysRevLett.103.105303}. Still there is no complete understanding why some model crystal systems are more convenient than others for particular FCI states. One way to resolve this problem would be to study  the underlying quantum geometry of the crystal system and its influence on the gap function of FCI in the scope of the usual approximation in the FQHE physics: the single mode approximation (SMA)~\cite{girvin-PhysRevB.33.2481}.

In practice finding favorable conditions for FCI means achieving the understanding of the interplay of several ingredients: The statistics of the underlying particles, the interactions, and the background lattice that produces the band with non-trivial topology characterized by a non-zero Chern number. We will focus our attention on the influence of the background lattice, and the structure of its particular band in which the FCI physics takes place. The band structure is characterized by two tensors: the Fubini-Study~(FS) (or quantum distance) metric and the Berry curvature.  They provide the description of the evolution of the quantum - mechanical state as it changes with the change of the Bloch momentum of the lattice. The question of the role of the FS metric in the context of FCI was first raised in Ref.~\cite{Roy-2012arXiv1208.2055R}. A recent work\cite{Neupert-PhysRevB.87.245103} showed how the FS metric is related to the current noise spectrum, a measurable quantity. While there is a general understanding that an almost flat Berry curvature favors the emergence of a FCI phase ~\cite{Parameswaran-2012PhRvB..85x1308P} (mimicking the constant magnetic field of FQHE), there is a partial understanding what is the influence of the FS metric. We know that there is no quantization of its value when integrated over the Bloch momentum phase space - Brillouin zone, like in the case of Berry curvature (that produces the Chern number). Quite generally its averaged value depends on the lattice parameters. Thus we cannot reduce the situation to the one of a single Landau level (FQHE) in which the relationship between, what we may identify as, Berry curvature and FS metric is fixed and not model dependent. Even on the mean field level (as we will show in our static structure factor (SSF) calculations in more detail) we may loose this relationship. In the geometrical picture of FQHE \cite{Haldane-PhysRevLett.107.116801} this relationship appears as a requirement for a unimodular  metric (a consequence of the fixed relationship between the applied flux and the number of particles). As pointed out in Ref.~\cite{Roy-2012arXiv1208.2055R}, an almost flat metric is needed to mimic the FQHE. Still there is a remaining question and an open problem that we will address in this work: how the system sustains the increase of the averaged FS metric from the FQHE value.

On the other hand we want to point out that there is the largely unexplored problem and question on the status of the SMA for FCI. In the case of FQHE, the SMA assumes a particular variational ansatz, based on the action of the projected to a fixed Landau level density operator on the ground state, in order to describe the first excited state .   In the search for favorable conditions for FCI, the SMA is the most likely tool that is available  if we want to understand the parameter (FS metric) dependence of the measure of the stability of the FCI state - its gap function. We consider the SMA in the context of FCI in the analogous manner as it is done in the FQHE; we assume that the first excited state is given by the action of the projected to a particular band density operator that acts on the ground state. In this work we will pay special attention to the SSF, the norm of the approximate state, although  ``f(k)" (the oscillator strength) function, i.e.  the expectation value of the Hamiltonian of the system in the variational ansatz for the density wave excited state is also important and needed. In the literature we find~\cite{Neupert-2012PhRvB..86c5125N} a general statement on the nature of the expansion of the f(k) function for FCI; the leading term in the long-distance expansion is quadratic in the low momentum due to the dependence of the Berry curvature on the Bloch momentum in FCI.
Because the gap function, in SMA, is the ratio of the f(k) function and the norm of the variational ansatz state i.e. SSF, this necessarily means that SSF has the same leading behavior. We want to point out that the derivation in the same reference~\cite{Neupert-2012PhRvB..86c5125N} was done with the usual averaging and long-wavelength limit procedures~\cite{Parameswaran-2012PhRvB..85x1308P,goerbig-2012epjb,Bernevig-2012PhysRevB.85.075128} in FCI, but which were not assumed or applied in the original work in Ref.~\cite{girvin-PhysRevB.33.2481} on the SMA in the context of FQHE. The behavior that was described in Ref.~\cite{girvin-PhysRevB.33.2481} is the leading behavior of the SSF and f(k) as quartic in small momenta. This is, we may say, a hallmark of the FQHE behavior.

In this paper we  will rederive this important result of Ref.~\cite{girvin-PhysRevB.33.2481} to set the stage for the discussion of the FCI case (Section~\ref{Subsection:SSFLaughlin}).  The quartic behavior in the FQHE context is a consequence of space symmetries and liquid, homogeneous nature of the system (Ref.~\cite{girvin-PhysRevB.33.2481} and
Section~\ref{Subsection:Discussion}). Here, in the introduction, we would like to point out that if FCI systems would remain characterized by the quartic behavior of the SSF that would mean that a description based on averaged quantities (of Berry curvature and metric) is sufficient to describe the long-distance behavior. At least in this domain one would not expect much difference between the FQHE and FCI behavior in their gap function and SSF. On the other hand with the quadratic behavior of SSF the role of the background variations would be very important in determining the gap function. FCIs would represent a non-trivial generalization of the FQHE behavior.

Thus it is natural to ask the following question: Whether, in a mean field picture at least we can talk about the scenario of Ref.~\cite{girvin-PhysRevB.33.2481} for FCI, or lower order terms in the SSF and f(k) that depend on the variations in the Berry curvature and FS metric will determine the gap function.
In this work we will study the SSF for FCI Laughlin-like states. In doing this one may resort to a numerical calculation of the SSF of FCI, which undoubtedly will shed most light on the nature of the expansion of the SSF - whether or not we have the quadratic term. Due to the approximations made and complicated expressions in Ref.~\cite{Neupert-2012PhRvB..86c5125N} that needs to be checked for concrete FCI states. On the other hand, as will be described in this work, one may build a mean field  picture (valid at least in the constant curvature and constant metric case)  based on analytical considerations in the long-distance limit. By analyzing the single particle physics of the band with non-trivial Chern number and its consequences on the SSF of FCI, we can find out that the effective (``plasmonic") density-density potential in the long-distance limit is inversely proportional to the averaged over Brillouin zone (BZ) FS metric. Thus the averaged over BZ FS metric plays the role of ``Landau level" metric in the framework of  the geometrical description of fractional quantum Hall systems~\cite{Haldane-PhysRevLett.107.116801} and an effective description of FCIs. In the effective description of the non-interacting CI band the density-density potential will ensure that the long-distance fluctuations are suppressed - that the non-interacting system is gapped, and the density is uniform in the long-distance limit. By applying the Feenberg formula~\cite{feenberg1969theory,Bares-PhysRevB.48.8636,Milovanovic-PhD} to the case of interacting quantum liquid we can track down the influence of the FS metric on the ensuing coefficient of the quartic term of the SSF. This will generalize the expression of Ref.~\cite{girvin-PhysRevB.33.2481}, formula (4.29) in the original work i.e. expression (\ref{final_eq}) below, to the FCI case. In this way we will address the open question: how the departure of the averaged value of FS metric from the FQHE value (i.e. departure from the unimodular metric description) affects the physics of the FCI state.

Thus the problem of calculating the SSF in the context of FCIs poses many difficulties, and makes transparent the very nature of the FCIs' complex background i.e. the Berry curvature and FS metric that vary over the BZ. The geometric picture of FQHE systems proposed in Ref.~\cite{Haldane-PhysRevLett.107.116801}, in the context of FCIs may find the most non-trivial realization where also extrinsic part of the metric degree of freedom (not the one related to interactions) is non-trivial. In this work, a definite answer for the SSF behavior in the long-distance limit will not be given, but, as we already announced, we will analyze the most relevant properties in the same limit by analytical means. It will be explained which conditions have to be fulfilled in order for the FQHE scenario (Ref.~\cite{girvin-PhysRevB.33.2481}) to occur in these systems. To illustrate our main conclusions we analyzed the underlying quantum geometry of the Haldane model~\cite{haldane-1988PhRvL..61.2015H} and the presence of bosonic FCI states in the phase diagram of the same model with on-site interaction (repulsion) only. We choose bosons in order to minimize the influence of the symmetry of the underlying lattice. We will present the whole phase diagram to have a better view of the influence of the underlying geometry and make comparison with our analytical predictions in the region where the magneto-phonon and the long-distance behavior should be relevant. We find that although the averaged FS metric departs strongly from the FQHE value (from the unimodular metric requirement) the FCI state persists for a while as both, our analytical and numerical work indicate.

The paper is organized as follows. In Section~\ref{Section:SSF} the derivation of SSF in the context of FQHE is reviewed. In Section~\ref{Section:FCISSPContribution} the role of the FS metric in the context of FCIs and FCI SSF is identified. In the scope of a mean field approximation together with the assumption of the Laughlin-Jastrow correlations, the long-distance limit of the SSF for Abelian FCI states is calculated in Section~\ref{Section:FCISSF}. The possibility for the same scenario of Ref.~\cite{girvin-PhysRevB.33.2481} in the context of FCI states is discussed in the same section. In Section~\ref{Section:FCIHaldane} the background degrees of freedom, the FS metric and Berry curvature, for the Haldane model based FCIs are analyzed. In the same section, Section~\ref{Section:FCIHaldane}, the phase diagram of the interacting system of bosons (that live on the lattice defined by the Haldane model) is presented. Based on the phase diagram, the identification of possible FCI regions and a comparison with the calculated background  properties and analytical results was made possible. Section~\ref{Section:Conclusions} is devoted to conclusions.

\section{The static structure factor and Laughlin case}\label{Section:SSF}
\subsection{Static structure factor}\label{Subsection:SSF}
We define static structure factor (SSF) as
\begin{equation}
s(q) = \frac{1}{V} \langle \Psi| \rho_{-q} \rho_q |\Psi \rangle, \label{def_SSF}
\end{equation}
where $V$ is the volume of the system, $ |\Psi \rangle $ is a normalized many-body wave function, and
\begin{equation}
\rho_q = \int \mathrm{d} \mathbf{r}\; \exp\{-\mathrm{i} \mathbf{q}\cdot \mathbf{r}\} \rho(\mathbf{r}),
\end{equation}
where
\begin{equation}
\rho(\mathbf{r}) = \sum_{i=1}^{N} \delta^{(2)}(\mathbf{r} - \mathbf{r}_i),
\end{equation}
is the density operator of the system of $N$ particles with coordinates $\{\mathbf{r}_i; i = 1, \ldots, N \}$. If we introduce the radial distribution function,
\begin{equation}
g(|\mathbf{r}_1 - \mathbf{r}_2|) = \frac{N (N-1)}{n^2} \int  \mathrm{d} \mathbf{r}_3 \cdots
\int  \mathrm{d} \mathbf{r}_N |\Psi( \mathbf{r}_1, \ldots, \mathbf{r}_N)|^2,
\end{equation}
where $n = N/V$ is the averaged density, we can rewrite Eq.~\ref{def_SSF} as
\begin{equation}
s(q) = n + n^2 \int \mathrm{d} \mathbf{r}\;  g(r) \exp\{-\mathrm{i} \mathbf{q} \mathbf{r}\} \label{defnz}
\end{equation}
We will call the first term in $s(q)$ single particle part and the second one two-body (correlation) part.

In the second quantization language with creation and annihilation operators in the coordinate space,
$\hat{\Psi}^{\dagger} (\mathbf{r})$ and $\hat{\Psi} (\mathbf{r})$, we have
\begin{equation}
[\hat{\Psi} (\mathbf{r}), \hat{\Psi}^{\dagger} (\mathbf{r}')]_{\pm} = \delta^{(2)}(\mathbf{r} -
\mathbf{r'}), \label{anticom_relations}
\end{equation}
where $+$ sign denotes a commutator in the case of bosons, and $-$ sign denotes an anticommutator in the case of fermions. In this language the corresponding expression for $s(q)$ is
\begin{eqnarray}
&& s(q) = \frac{1}{V} \langle \Psi| \int \mathrm{d} \mathbf{r} \; \hat{\Psi}^{\dagger} (\mathbf{r})\hat{\Psi} (\mathbf{r}) \;| \Psi \rangle \nonumber \\
&& + \frac{1}{V}\int \mathrm{d} \mathbf{r} \int \mathrm{d} \mathbf{r'} \; \langle \Psi| \hat{\Psi}^{\dagger} (\mathbf{r}) \hat{\Psi}^{\dagger}(\mathbf{r'}) \hat{\Psi} (\mathbf{r'})  \hat{\Psi} (\mathbf{r}) |\Psi \rangle \times \nonumber \\
&&\exp\{-\mathrm{i} \mathbf{q}(\mathbf{r}- \mathbf{r'})\}  .\label{secondq}
\end{eqnarray}
In other words, by using the basic algebra of particle operators i.e.  (anti)commuting relation in Eq.~\ref{anticom_relations} we can recover separate single part (first term of Eq.~\ref{secondq}) and the two-body part (second term of Eq.~\ref{secondq}).

Due to the Dirac delta functions in Eq.(\ref{anticom_relations}) in this what we may call an unprojected case, the single particle part is trivial, and equal to the value of the density. In the following, in the context of the lowest Landau level (LLL) physics, instead of the delta functions we will have Gaussians.  This will lead to a non-trivial single particle part of the SSF defined in the
LLL by using density operators projected to the LLL.
\subsection{Quantum Hall case - Laughlin case}\label{Subsection:SSFLaughlin}
We would like to calculate the same quantity SSF, defined in Eq.~\ref{def_SSF} where $ |\Psi \rangle $ is the normalized Laughlin wave function. With respect to Ref.~\cite{girvin-PhysRevB.33.2481}, we use the definition of SSF given by Eq.~\ref{def_SSF}, in which we divide the density-density correlator by volume ($V$) instead of the number of particles ($N$). The definition used in Ref.~\cite{girvin-PhysRevB.33.2481} corresponds to the norm of the model state used in SMA. In the following we will summarize the results of weak coupling plasma approach as described in Refs.~\cite{Milovanovic-PhD} and~\cite{Milovanovic-PhysRevB.59.10757}. We will rederive the small-momentum behavior of SSF found in Ref.~\cite{girvin-PhysRevB.33.2481} in the manner (i.e. using the weak coupling approach and the division into the single and two-particle part) that is convenient for the later application in the FCI case in Sections~\ref{Section:FCISSPContribution} and \ref{Section:FCISSF}.

If we assume weak coupling in the Laughlin plasma approach~\cite{laughlin83prl1395,Laughlin-Prange} we can study expansions of expectation values as in Eq.~\ref{def_SSF} in terms of the 2D Coulomb plasma interaction,
\begin{equation}
v(q) = - \frac{4 \pi m}{|q|^2},
\end{equation}
where positive integer $m$ is connected to the filling factor $\nu$ of the quantum Hall system as $\nu = 1/m$. Because of the expected screening of the 2D plasma, the contributions in the weak coupling perturbative approach (in this unprojected case with the usual density operators) can be organized as in
\begin{equation}
s(q) = \frac{s_{0}(q)}{1 - v(q) s_{0}(q)}, \label{geo_sum}
\end{equation}
where $s_{0}(k)$ represents the contribution from ``irreducible diagrams". In the lowest order in $m$, $s_{0}(q) \approx n = 1/2 \pi m$. We set the magnetic length to one, $l_B \equiv 1$.

In the lowest Landau level (LLL) we can define the projected density,
\begin{equation}
\tilde{\rho}_{q} = \int \mathrm{d}^2 z \Psi^{\dagger}(z) \exp\{\mathrm{i} q \frac{\partial}{\partial z}\} \exp\{ \mathrm{i}  \frac{q^{*} z}{2} \}\Psi(z), \label{prodensity}
\end{equation}
where $\Psi(z)$ are second quantized operators in the LLL,
\begin{equation}
\Psi = \sum_{l = 0}^\infty \hat{a}_l \frac{1}{\sqrt{2\pi 2^l l!}} z^l \exp\{- \frac{1}{4} |z|^2\},
\end{equation}
$z = x + \mathrm{i} y$ is the complex 2D coordinate and $[\hat{a}_l, \hat{a}_m^{\dagger}]_\pm = \delta_{l,m}$. The derivatives in Eq.~\ref{prodensity} act only on the holomorphic part (dependent only on $z$) of $\Psi(z)$. We have
\begin{eqnarray}
&&\tilde{\rho}_{-q} \tilde{\rho}_{q} = \nonumber \\
&&\int \mathrm{d}^2 z \int \mathrm{d}^2 z'
[\exp\{- \mathrm{i} q^{*} \frac{\partial}{\partial z^{'*}}\}\exp\{- \mathrm{i} \frac{q z^{'*}}{2} \}
\Psi^{\dagger}(z')] \Psi(z') \nonumber \\
&& \times \Psi^{\dagger}(z) \exp\{\mathrm{i} q \frac{\partial}{\partial z}\} \exp\{ \mathrm{i}  \frac{q^{*} z}{2} \} \Psi(z), \label{lll_case}
\end{eqnarray}
Similarly to the unprojected case, Eqs.~\ref{anticom_relations} and~\ref{secondq},  we can use the (anti)commutation relations of projected single particle operators, $\Psi(z)$, to get the division in the single and two-body part of the SSF defined by these projected density operators. As we already discussed the single particle part of a SSF corresponds to the diagonal contribution either in momentum or coordinate space that refers to a single particle. We will denote by
$\langle \tilde{\rho}_{-q} \tilde{\rho}_{q} \rangle|_{\text{single}}$, the single part of the projected SSF.
Due to
\begin{eqnarray}
[\Psi(z'), \Psi^{\dagger}(z)]_\pm = \frac{1}{2 \pi} \exp\{\frac{z^* z'}{2}\}
\exp\{-\frac{|z|^2}{4}\} \exp\{-\frac{|z'|^2}{4}\}, \nonumber \\
\end{eqnarray}
i.e. equality of the (anti)commutator to the LLL delta function, the single particle part of the projected SSF is
\begin{equation}
\langle \tilde{\rho}_{-q} \tilde{\rho}_{q} \rangle|_{\text{single}} =
n \exp\{- \frac{|q|^2}{2}\}. \label{singleSSF}
\end{equation}
The two-particle correlations (i.e. when $z' \neq z$) stay the same (examine Eqs.~\ref{lll_case} and~\ref{secondq} with Laughlin wave function in that case) and therefore the expression for the projected SSF is
\begin{equation}
\tilde{s}(q) = n \exp\{- \frac{|q|^2}{2}\} + \frac{s_{0}(q)}{1 - v(q) s_{0}(q)}  - n . \label{pSSF}
\end{equation}
The important correction to the $s_{0}(q) \approx n$ approximation i.e. the next contribution to the sum of irreducible parts, $s_{0}(q)$, of order $m^0$, which we will denote by $\delta_q$, is made of a bubble diagram. In the diagram the two interaction lines are screened. Therefore the contribution is
\begin{equation}
\delta_q = n^2 \frac{1}{2} \int \frac{\mathrm{d}^2k}{(2\pi)^2} V_{\text{eff}}(\mathbf{q}-\mathbf{k}) V_{\text{eff}}(\mathbf{k}),
\label{first_eq}
\end{equation}
where $1/2$ is a symmetry factor and
\begin{equation}
V_{\text{eff}}(\mathbf{q}) = \frac{V(\mathbf{q})}{1 - n V(\mathbf{q})} = - \frac{4 \pi m}{|\mathbf{q}|^2 + 2}.
\end{equation}
The correction for $\mathbf{q} = \mathbf{0}$ is
\begin{equation}
\delta_0 = n \frac{m}{2}.
\end{equation}
We checked that the correction of the order $m^0$ equal to $\delta_q$ in $s_0(q) \approx n + \delta_q$ will not change the resulting behavior due to the inclusion of $s_0(q) \approx n + \delta_0$ in Eq.~\ref{pSSF} in the weak coupling case:
\begin{equation}
\tilde{s}(q) \approx \frac{(m - 1)}{8} |q|^4 n. \label{final_eq}
\end{equation}
If we assume that the analytical continuation for $m \geq 1$ is valid we recovered (up to the density difference in the normalization) the formula of Ref.~\cite{girvin-PhysRevB.33.2481}, expression (4.29) in the original work, that follows from the 2D plasma compressibility rule.

\section{FCI case - single particle contribution}\label{Section:FCISSPContribution}
We start with a lattice system in which Bloch Hamiltonian $H(p)$ is diagonal with eigenvalues, $\epsilon_m(p)$, where $m = 1,\ldots,r$ denotes the band index, and eigenstates, $u^{m}_{\alpha, p}$, where $\alpha =1,\ldots,r$ denotes the orbital index in a unite cell, and $p$ is the Bloch momentum, i.e.
\begin{equation}
\sum_{\beta} h_{\alpha, \beta}(p) u_{\beta,p}^{m} = \epsilon_m(p) u_{\alpha,p}^{m}.
\end{equation}
In order to study the partially filled band with Chern number equal to one in which a FCI state occurs, we confine our description to that band, and drop the index $m$ in the following. In the case of FCI we take the projected density to a single band to be  defined as in Ref.~\cite{Bernevig-2012PhysRevB.85.075128} i.e.
\begin{equation}
\tilde{\rho}_q = \sum_p u_{\alpha,p}^{*} u_{\alpha,p+q} \gamma^{\dagger}_{p}
\gamma_{p+q}, \label{fci_density}
\end{equation}
where the summation on the repeated orbital Greek index $(\alpha)$ is assumed, $p$ and $q$ are Bloch momenta, and $\gamma_p$ are normal mode operators.

Therefore
\begin{eqnarray}
&&\tilde{\rho}_{-q} \tilde{\rho}_{q}= \nonumber \\
&&\sum_{p_1,p_2} u_{\alpha,p_1}^{*} u_{\alpha,p_1-q} u_{\alpha,p_2}^{*} u_{\alpha,p_2+q} \times \nonumber \\
&&\gamma^{\dagger}_{p_1}
\gamma_{p_{1}-q}\gamma^{\dagger}_{p_2}
\gamma_{p_{2}+q},
\end{eqnarray}
and because
\begin{equation}
[\gamma_{p_{1}-q},\gamma^{\dagger}_{p_2}]_\pm = \delta_{p_{1}-q,p_2},
\end{equation}
we have
\begin{eqnarray}
&&\langle \tilde{\rho}_{-q} \tilde{\rho}_{q}\rangle|_{\text{single}}= \nonumber \\
&&\sum_{p} u_{\alpha_1,p}^{*} u_{\alpha_1,p-q} u_{\alpha_2,p-q}^{*} u_{\alpha_2,p}\; n_p.
\label{no_expansion_single_part}
\end{eqnarray}
 In this expression, $n_p \equiv \langle \gamma^{\dagger}_{p} \gamma_{p}\rangle$ is the occupation of the Bloch momentum $p$ in the many-body FCI state.

After a few steps, which are described in Appendix A, the expansion in small momentum $q$ of the single part of the unprojected SSF for the FCI state is
\begin{equation}
\langle \tilde{\rho}_{-q} \tilde{\rho}_{q}\rangle |_{\text{single}} = n - q_i q_j n \overline{g_{ij}^{FS}} + \frac{q_i q_j q_k}{2} n  \overline{\partial_k g_{ij}^{FS}}  + o(q^4), \label{expansion}
\end{equation}
where we assumed summations over repeated indices. With the Fubini Study metric defined by,
\begin{eqnarray}
&& g_{ij}^{FS}(p) = \frac{1}{2}  [ \partial_i u_{\alpha_1,p} \partial_j u_{\alpha_1,p}^{*} + \partial_j u_{\alpha_1,p} \partial_i u_{\alpha_1,p}^{*} -\nonumber \\
&& \partial_i u_{\alpha_1,p} u_{\alpha_1,p}^{*} u_{\alpha_2,p}\partial_j u_{\alpha_2,p}^{*} - \partial_j u_{\alpha_1,p} u_{\alpha_1,p}^{*} u_{\alpha_2,p}\partial_i u_{\alpha_2,p}^{*}] \nonumber \\
\end{eqnarray}
the coefficients,
\begin{equation}
\overline{g_{ij}^{FS}} = \frac{\sum_p g_{ij}^{FS}(p) n_p}{n},
\end{equation}
and
\begin{equation}
\overline{\partial_k g_{ij}^{FS}} = \frac{\sum_p \partial_k g_{ij}^{FS}(p) n_p}{n}, \label{derivative}
\end{equation}
are averages over the whole BZ. If we assume $n_p = \text{constant}$, which we expect to hold at least approximately in the FCI state, see  Section~\ref{Subsection:Discussion} (Eq.~\ref{liquid_condition}) for an explanation of this point,  we can substitute the averages over occupation number to the ones over Brillouin zone. Because of the periodicity in the $k$-space we have that the averaged derivatives over the metric (Eq.~\ref{derivative}) are equal to zero in the case of FCIs. Also we can argue that, assuming the inversion symmetry, the expression in Eq.~\ref{derivative} is zero. Thus
\begin{equation}
\langle \tilde{\rho}_{-q} \tilde{\rho}_{q}\rangle |_{\text{single}} = n - q_i q_j n \overline{g_{ij}^{FS}}  + o(q^4). \label{expansion2}
\end{equation}
Comparing with the Laughlin case,  Eqs.~\ref{singleSSF} and~\ref{pSSF}, we expect that the quadratic term corresponds to the plasmonic part that has to be canceled in the projected (to the band) SSF if  the FQHE scenario occurs in the context of FCI.

Continuing the analysis of the single part to the quartic order we have
\begin{eqnarray}
\langle \tilde{\rho}_{-q} \tilde{\rho}_{q}\rangle |_{\text{single}} =&& n - q_i q_j n\; \overline{g_{ij}^{FS}}
+ q_i q_j q_k q_l n\; \overline{h_{ijkl}} \nonumber \\
&&+ \; \text{higher order terms}, \label{expansion3}
\end{eqnarray}
where
\begin{eqnarray}\label{Ehvalue}
h_{ijkl} & = & \frac{1}{4!} [ u_{\alpha,p} \partial_k \partial_i \partial_j \partial_l u_{\alpha,p}^{*} +
\partial_k \partial_i \partial_j \partial_l u_{\alpha,p}\; u_{\alpha,p}^{*} \nonumber \\
&& + 4 \partial_i u_{\alpha_1,p} u_{\alpha_1,p}^{*}\; u_{\alpha_2,p}\partial_k \partial_j \partial_l u_{\alpha_2,p}^{*}+ \nonumber \\
&&+  4\; \partial_i \partial_k \partial_l u_{\alpha_1,p} u_{\alpha_1,p}^{*}\; u_{\alpha_2,p} \partial_j u_{\alpha_2,p}^{*}+ \nonumber \\
&&+ 6\; \partial_i \partial_k  u_{\alpha_1,p} u_{\alpha_1,p}^{*}\; u_{\alpha_2,p} \partial_j \partial_l u_{\alpha_2,p}^{*}],
\end{eqnarray}
and
\begin{equation}\label{Eh}
\overline{h_{ijkl}} = \frac{\sum_p h_{ijkl}(p) n_p}{n}.
\end{equation}

In the analogy with the Laughlin case we would expect that the coefficient of the quartic term is a ``square of metric", but the tensor $h_{ijkl}$ can not be greatly simplified without further assumptions. Assuming $n_p = \text{constant}$, we can shift the sum over $p$ to the one over $P = p - q/2$, and discuss the product, $u_{\alpha, P + q/2}^{*} u_{\alpha, P - q/2}$ separately. In this case the quartic coefficient becomes,
\begin{equation}
\overline{h_{ijkl}^{FCI}} = \frac{\overline{g_{ij}^{FS}g_{ij}^{FS}}}{4} + \frac{\overline{g_{ijkl}}}{4}, \label{hFCI}
\end{equation}
where $g_{ijkl}$ is a tensor. By analyzing $ u_{\alpha, P + q/2}^{*} u_{\alpha, P - q/2}$ product and rewriting it as
\begin{equation}
u_{\alpha, P + \frac{q}{2}}^{*} u_{\alpha, P - \frac{q}{2}} = \exp\{-\mathrm{i}q_i A_i\} f(u, u^{*}),
\end{equation}
where $ A_i = - i u^{*}_{\alpha, p} \partial u_{\alpha, p}$, the Berry connection, we can find out the expression for the $g_{ijkl}$ tensor in its gauge invariant form,
\begin{eqnarray}
2 g_{ijkl} & = & - \frac{1}{3} ( \partial_i u_{\alpha,p} \partial_j \partial_k \partial_l u_{\alpha,p}^{*} +
 \partial_i \partial_j \partial_k u_{\alpha,p}\; \partial_l u_{\alpha,p}^{*}) \nonumber \\
&& + \frac{1}{3} i A_i (  u_{\alpha,p} \partial_j \partial_k \partial_l  u_{\alpha,p}^{*} -
  \partial_i \partial_k \partial_l u_{\alpha,p} u_{\alpha,p}^{*} \nonumber \\
&& +  \partial_i \partial_k  u_{\alpha,p} \partial_j u_{\alpha,p}^{*} - \partial_i u_{\alpha,p} \partial_j \partial_l u_{\alpha,p}^{*}) \nonumber \\
&& + 4 A_i A_j \partial_k u_{\alpha,p}\partial_l u_{\alpha,p}^{*} - 2 A_i A_j A_k A_l.
\end{eqnarray}
Even the condition on the constancy of the metric, $ \partial_i \partial_j g_{kl}^{FS} = 0$, does not simplify the form of the $g_{ijkl}$ tensor and the coefficient $h_{ijkl}^{FCI}$. If we choose the LLL basis of Ref.~\onlinecite{wu110prl106802}, for which $ g_{11}^{FS} =  g_{22}^{FS} = \overline{B}/2$, where $\overline{B}$ is the averaged Berry curvature, i.e.
\begin{equation}
\overline{B} = \frac{\sum_p B}{A_{\text{BZ}}}= \frac{2\pi C}{A_{\text{BZ}}},
\end{equation}
where $A_{\text{BZ}}$ is the area of the BZ, and $ g_{12}^{FS} =  g_{21}^{FS} = 0$, we expect to recover the ``square of metric" form of the coefficient. Otherwise, for a general FCI, we expect Eq.~\ref{hFCI} to hold.

The main result of this Section is Eq.(\ref{expansion2}) in which by comparing to the FQHE expression (\ref{singleSSF}), we can identify the averaged over BZ FS metric as the relevant metric in the long-distance limit. With respect to the metric defined in the context of the geometrical picture of (continuum) FQHE , Ref.~\cite{Haldane-PhysRevLett.107.116801}, our use of the FS metric brings factor of two, $ g_{ij}^{FQHE} = 2 g_{ij}^{FS}$, when we relate them. Thus the requirement for the unimodular metric in the context of FQHE translates to the determinant of the FS metric being equal to 1/4 in appropriate units. More on this relationship can be found in Section \ref{Subsection:Discussion}.
-----------------------------------
\section{SSF for FCI}\label{Section:FCISSF}
\subsection{A mean field picture}\label{Subsection:Meanfield}
In the following we will consider a  possibility that the absence of the quadratic term in the expansion of the projected SSF in the FQHE also occurs in the context of the FCI physics. In a mean field picture we may expect that the FCI system {\em in the long-wavelength limit} is a system with density $ n = 1/2 \pi m l_B^2$, fixed by the value of $l_B^2 \equiv \overline{B}$~\cite{Parameswaran-2012PhRvB..85x1308P} (i.e. averaged Berry curvature i.e. Chern number), and  that two body correlations are described in the same limit with an effective long-range density-density potential,
\begin{equation}
v^{FCI}_{MF}(q) = - \frac{2 \pi m}{q_i q_j  \overline{g_{ij}^{FS}}}. \label{mfint}
\end{equation}
The form of the effective long-range density-density potential is fixed by the assumption that the cancelation occurs.

The assumption we made is that in an averaged picture the two body correlations are still Laughlin-like with a constant metric though in the FCI case we have the single-particle properties, Berry curvature and metric that vary with the Bloch momentum.
In the following we will consider that
\begin{eqnarray}
&& \overline{g_{11}^{FS}} =  \overline{g_{22}^{FS}} = g \geq \frac{\overline{B}}{2} \;\; \text{and} \nonumber \\
&&\overline{g_{12}^{FS}} =  \overline{g_{21}^{FS}} = 0,
\end{eqnarray}
i.e. the averaged over BZ FS metric is diagonal with diagonal element equal to $g$. The value of $g$ has the lower bound, $\bar{B}/2$, as explained in Ref.~\cite{Roy-2012arXiv1208.2055R}. The lower bound, $\bar{B}/2$, corresponds to the FQHE case (``unimodular metric" if metric and Berry curvature are constant) as we will explain in Section \ref{Subsection:Discussion}. The diagonal form of the averaged metric will hold in the context of FCI states based on the Haldane model Chern insulator, when the assumption $n_p = \text{constant}$ is applied.

When calculating the SSF for FCI in the approximation we adopted, we can make one of the two following assumptions. We can suppose that, in the long-wavelength limit, the system is described by the Laughlin wave function (or its generalizations with in general ``incongruent" relationship between flux and particle positions or unknown short distance behavior). In that case, we apply the plasma approach of Section~\ref{Section:SSF}, considering only the long-wavelength domain. Equivalently we can assume that we have a bosonic quantum liquid system with the long-range potential to which we can apply the Feenberg formula~\cite{feenberg1969theory,Bares-PhysRevB.48.8636} (Eq.~\ref{geo_sum}) in this limit.

Either way, assuming the analyticity in $m$  and repeating the steps of Eqs.~\ref{first_eq}-\ref{final_eq}, we find that the two particle part of the SSF of FCI in this limit behaves as
\begin{eqnarray}
&& \lim_{q \rightarrow 0} \frac{s_{0}^{FCI}(q)}{1 - v^{FCI}_{MF}(q) s_{0}^{FCI}(q)} \rightarrow \nonumber \\
&& g \; n |q|^2 +  g^2 \; n |q|^4 ( \frac{m}{4} \frac{\overline{B}}{g} - 1).
\end{eqnarray}
Within the assumptions made (and that we work with the FS metric, $ \overline{g_{11}^{FS}} =  \overline{g_{22}^{FS}} = g$) we can conclude that the form of the projected to a band SSF for a FCI state is
\begin{eqnarray}
\tilde{s}^{FCI}(q) =&&  \frac{g^2 \; n |q|^4}{2} ( m \frac{\overline{B}}{2g} - 2)\nonumber \\
&& + n \; q_i q_j q_k q_l  \frac{\overline{g_{ij}^{FS}g_{kl}^{FS}}+ \overline{g_{ijkl}}}{4}. \label{final_quartic}
\end{eqnarray}
In the case $g = \overline{B}/2$, when the Berry curvature and metric are constant, we expect to recover the usual QHE form (expression (\ref{final_eq})).

\subsection{Discussion}\label{Subsection:Discussion}
According to the Ref.~\cite{Neupert-2012PhRvB..86c5125N} the Berry curvature that varies over the BZ will produce the quadratic term in the expansion of the $f_k$ function (Ref.~\cite{girvin-PhysRevB.33.2481}, expression (4.12)) in the SMA for the FCI. The authors concluded that necessarily the projected SSF of the FCI has to have the leading  quadratic term in order to have a finite gap in the SMA.

We applied a mean field approach in calculating two-body correlations for the SSF of FCI. It is likely that only an exact numerical calculation based on a concrete FCI state may determine whether the quadratic term in the expansion of the SSF is present or that the quartic term is dominant in determining the physics and energetics of FCIs (as in the usual FQHE case).

If we nevertheless maintain that in a mean field picture the formula~(\ref{final_quartic}) enters the expression for the gap function (as a denominator - a norm of the SMA state) in the SMA of the FCI state, we can conclude that large $g \gg \overline{B}/2$ may induce an instability. In other words $\tilde{s}^{FCI}(q)$ will become negative, which cannot be true for a positive definite quantity, and this would signal an instability towards a gapless state ($\tilde{s}^{FCI}(q) \sim q^3$). In reality we might expect that either large discrepancy between $g$ and the lower bound $\overline{B}/2$, or strong fluctuations of the FCI metric may lead to a gapless state. In order to investigate this question we calculated $\overline{g_{ij}^{FS}}$, $\overline{h_{ijkl}}$, and the standard deviation of the FS metric from its averaged value, $\overline{g_{ij}^{FS}}$, in the Brillouin zone for a particular model. We will present this in Section~\ref{Section:FCIHaldane}.

To understand better (see also~\cite{Haldane-2011arXiv1112.0990H,bernevig-PhysRevLett.100.246802,Yang-PhysRevB.85.165318}) the absence of the quadratic term in the FQHE we will discuss the case $ g = \overline{B}/2$ (diagonal and constant metric) and $ B (\text{Berry curvature}) = \bar{B}$ in the FCI context. Expanding the expression in Eq.~\ref{fci_density} for the projected density to a single band we have for the linear term in $q$:
\begin{equation}
\tilde{\rho}_{q}|_{\text{linear}} = q^k \sum_p
\{\mathrm{i} A_{k}(p) \gamma_p^{\dagger} \gamma_p + \gamma_p^{\dagger} \frac{\partial}{\partial p_k} \gamma_p \}  \equiv q^k T_k . \label{linear}
\end{equation}
In the first quantization picture the operator $T_k$ is
\begin{equation}
T_k = \sum_{i=1}^{N} \{\mathrm{i} A_{k}(p_i) + \frac{\partial}{\partial p^k_i}\}, \;\;\; k = x,y
\end{equation}
Using the complex representation we can rewrite the linear term in Eq.~\ref{linear} in the radial gauge as
\begin{equation}
\sum_{i=1}^{N} \{q (\frac{\bar{B}}{4} p_i^{*} + \frac{\partial}{\partial p_i}) + q^{*} (- \frac{\bar{B}}{4} p_i + \frac{\partial}{\partial p_i^{*}})\}.
\end{equation}
The solution must be of the form,
\begin{equation}
\Psi_0 = f(\{p_i\}) \exp\{- \frac{1}{4} \sum |p_i|^2\},
\end{equation}
i.e. belong to the LLL; the operators, $R_i = (\overline{B}/4) p_i^{*} + \partial/\partial p_i$ and $R_i^{\dagger} = (\overline{B}/4) p_i - \partial/\partial p_i^{*}$, we recognize, corresponding to guiding center coordinates in the momentum representation of the QH problem. They make the simple bosonic algebra, $[a_i, a_i^{\dagger}] = 1$, if we take $\bar{B} = 1$ and $ a_i = \sqrt{2} R_i$ and $ a_i^{\dagger} = \sqrt{2} R_i^{\dagger}$ of the LLL for each particle. In this representation  SSF can be expressed as
\begin{eqnarray}
&&\sum_{i,j} \langle :\exp\{q^{*} R_i^{\dagger} - q  R_i\}: :\exp\{- q^{*} R_j^{\dagger} + q  R_j\}: \rangle -
\nonumber \\
&& \sum_{i} \langle :\exp\{q^{*} R_i^{\dagger} - q  R_i\}: \rangle \sum_{j}\langle :\exp\{- q^{*} R_j^{\dagger} + q  R_j\}: \rangle , \label{no_def} \nonumber \\
&&
\end{eqnarray}
where : : sign denotes the normal ordering. This definition implies the usage of the density operators in the SSF calculation that differs by the factor $ \exp\{ - |q|^2/2\}$ from the usual~\cite{girvin-PhysRevB.33.2481} operators. Nevertheless, this trivial difference should not affect the absence of the quadratic term. Thus applying the expression in Eq.~\ref{no_def}, we see that, after a ground state value subtraction i.e. normal ordering - see Appendix for explanation of this point, the quadratic term will not exist if and only if
\begin{equation}
\sum_{i} R_i \Psi_0 = 0 . \label{lcondition}
\end{equation}
We know that this is satisfied in the disk and spherical geometry of a continuum system~\cite{bernevig-PhysRevLett.100.246802,Moore1991362}, and expect to hold even in the lattice system. The main reason for this is that the generator of translation should annihilate the ground state which is a homogenous, liquid state.

Therefore it is the existence of a homogenous ground state that is annihilated by the translation generator plus the existence of the bosonic algebra of the LLL that ensures the absence of the quadratic term, and appearance of the leading quartic term in the FQHE.

If we believe that the same scenario will happen in FCI we may consider the possibility that locally, in the BZ, even for varying curvature, we can have the bosonic algebra;
\begin{equation}
R_i = \frac{\bar{B}+ \delta B}{4} p_i^{*} + \frac{\partial}{\partial p_i} \;\text{and}\;
R_i^{\dagger} = \frac{\bar{B} + \delta B}{4} p_i - \frac{\partial}{\partial p_i^{*}},
\end{equation}
where $\delta B$ is a weakly dependent function on $\mathbf{p}$. With the condition that the FCI state must satisfy,
\begin{equation}
\sum_{i} R_i \Psi_{\text{FCI}}(\{\mathbf{p}_i\}) = 0 , \label{liquid_condition}
\end{equation}
and the normal ordering prescription for the bosonic algebra at each $\mathbf{p}$, the quadratic term will be absent in the low-momentum SSF expansion.

As we will see in a particular example of a bosonic FCI state based on the Haldane model
at and around the point, $\mathbf{p} = (0,0)$ in the BZ, which is a long-distance expansion point, in this particular model,  the Berry curvature is zero.  Thus the effective form of the ground state in this long-wavelength limit (when the particle momenta are small) may be of the Jastrow-Laughlin  form, i.e.
\begin{equation}
\prod_{i<j} {|p_i - p_j|^{\gamma}},
\end{equation}
where $\gamma$ is a constant. This form will
 satisfy Eq.~\ref{liquid_condition} in the limit for which the Berry curvature and metric matrix elements, see Figs.~\ref{particularG} and~\ref{particularB} below, are zero. With the assumption that the cancelation of the quadratic terms in the small momentum expansion of the projected SSF occurs we expect $\gamma = \frac{m \overline{B}}{2 g}$.

In this subsection we provided arguments why Girvin-MacDonald-Platzman scenario may be relevant in the FCI context, and in the following, when a comparison with numerical results are made, we will use the expression (\ref{final_quartic})  as the description of the leading behavior of the projected SSF.

\section{FCIs based on the Haldane model}\label{Section:FCIHaldane}
\subsection{Exact diagonalization results}
\begin{figure}[h]
\centering
\includegraphics[width=0.6\linewidth]{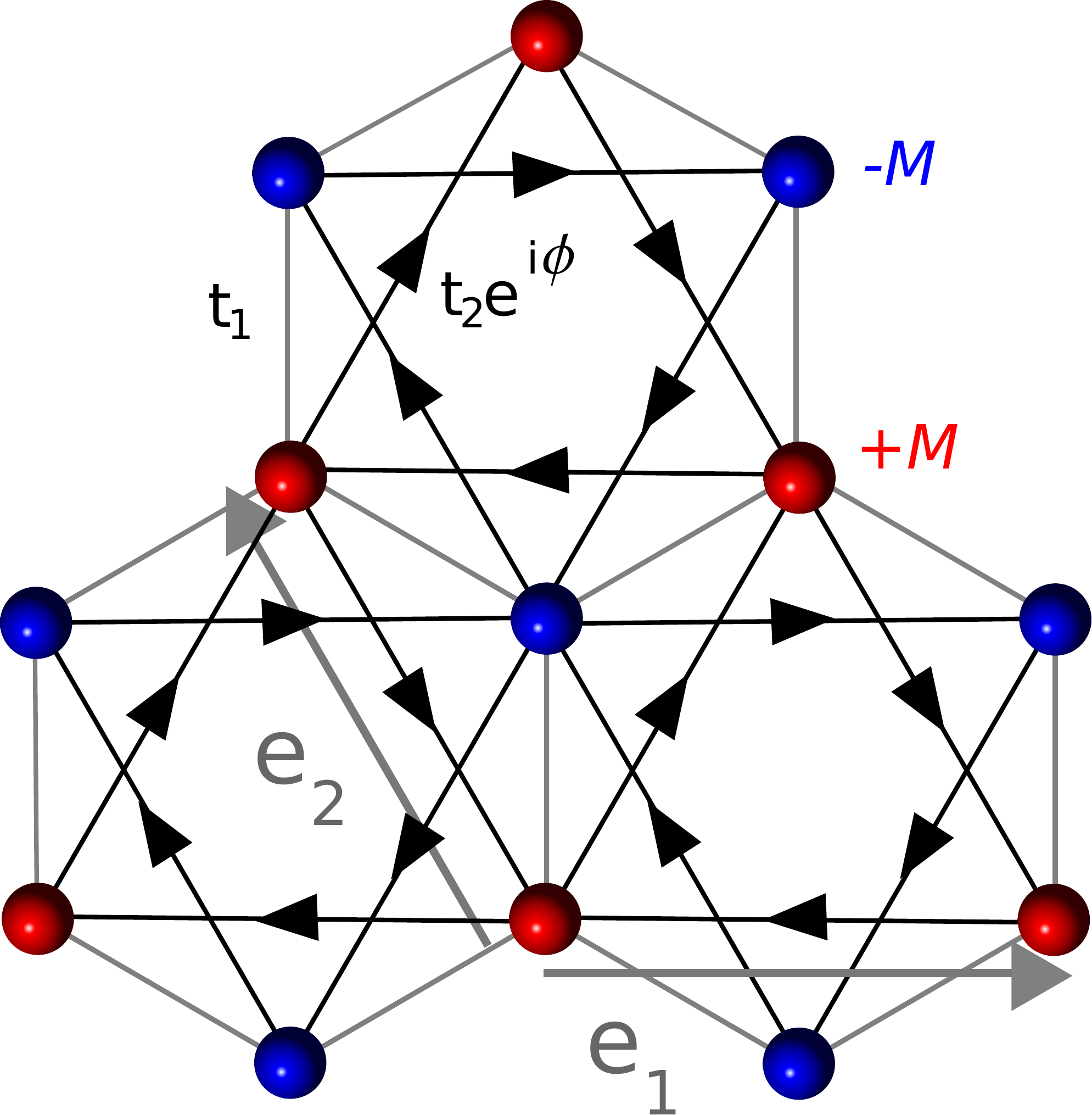}
\caption{\label{haldane} The Haldane model on the honeycomb lattice with $A$ (in red)
and $B$ (in blue) sublattices. The lattice translation vectors are $\mathbf{e}_1$ and $\mathbf{e}_2$. The amplitude of the nearest neighbor hopping is $t_1$ and the next nearest neighbor hopping is $t_2\exp(\mathrm{i}\phi)$ (in the direction of arrows). The sublattice chemical potential is set to $+M$ on $A$ sites and $-M$ on $B$ sites.
}
\end{figure}

The Haldane honeycomb model~\cite{haldane-1988PhRvL..61.2015H} is the first studied example of a
Chern insulator. Several numerical evidences of a robust FCI have been reported for bosons with on-site repulsion on such a lattice~\cite{wang-PhysRevLett.107.146803,wang-PhysRevLett.108.126805,Scaffidi-PhysRevLett.109.246805}. Weaker FCIs have also been observed when bosons are replaced by fermions~\cite{Wu-2012PhysRevB.85.075116} (Note that the on-site interaction is then replaced by a nearest neighbor interaction). We will use the honeycomb lattice layout of Ref.~\cite{neupert-PhysRevLett.106.236804}, as shown in Fig.~\ref{haldane}. The one-body Hamiltonian can be written in Bloch form as
$h(\mathbf{k})=d_0\mathbb{I}+\sum_i d_i\sigma_i$ using the Pauli matrices and where
\begin{eqnarray}
d_0&=&2t_2\cos\phi \; \left(\cos k_x + \cos k_y + \cos(k_x+k_y) \right),\nonumber\\
d_x&=&t_1 \left(1+\cos(k_x+k_y)+\cos k_y\right),\label{blochhaldane}\\
d_y&=&-t_1 \left(\sin(k_x+k_y)+\sin k_y\right),\nonumber\\
d_z&=&M\;+\;2t_2\sin\phi \;\left(\sin k_x + \sin k_y - \sin(k_x+k_y)\right).\nonumber
\end{eqnarray}
$M$ (resp. $-M$) is the chemical potential added to the $A$ (resp. $B$) sites, $t_1$ is the amplitude of the (real) nearest neighbor hopping term and $t_2 \exp (\mathrm{i} \phi)$ the complex amplitude if the next nearest neighbor hopping term. The two components of the lattice momentum $\mathbf{k}$ are defined as $k_x=\mathbf{k}\cdot \mathbf{e}_1$ and $k_y=\mathbf{k}\cdot \mathbf{e}_2$, where $\mathbf{e_1}$ and $\mathbf{e_2}$ are the lattice vectors. For our numerical calculations, we set $t_1=t_2$. The Haldane model has two bands. If $M/t_1 > 3 \sqrt{3} \sin (\phi)$, the two bands are trivial. If $M/t_1 < 3 \sqrt{3} \sin (\phi)$ then each band carries a non-zero Chern number (either $C=+1$ or $C=-1$).

We consider $N$ bosons on the Haldane honeycomb model with a lattice of $N_x$ unit cells in the $\mathbf{e}_1$ direction and $N_y$ unit cells in the $\mathbf{e}_2$ direction. The filling factor is thus defined as $\nu=N/(N_x\cdot N_y)$. We add on-site Hubbard-type density-density interaction $H_{\rm int} = \sum_{i} : n_i n_i:$, where the sum runs over all the sites. To focus on the band topological properties, we use the flat-band approximation described in Ref.~\cite{regnault-PhysRevX.1.021014}: We start from the original Bloch Hamiltonian $h(\mathbf{k})=\sum_{\alpha =1}^{2} E_\alpha(\mathbf{k}) {\cal P}_\alpha(\mathbf{k})$ where $E_\alpha(\mathbf{k})$ and ${\cal P}_\alpha(\mathbf{k})$ are the dispersion  and the projector onto the $\alpha$-th band, respectively. Then we focus on the lowest band and consider the effective one-body flat band Hamiltonian $h_{\rm eff}(\mathbf{k})= {\cal P}_1(\mathbf{k})$. From a physical perspective, it means that we set the band gap to infinity and we make the lowest band completely flat. In this approximation, the effective many-body hamiltonian writes $H_{\rm eff}= {\cal P}_1 H_{\rm int} {\cal P}_1$.

To study the stability of the FCI phase, we focus on the energy spectrum. In the FCI regime at filling factor $\nu=1/2$, the Laughlin-like state on a torus geometry is characterized by two almost degenerate low energy states separated by a large from higher energy excitations. A typical low energy spectrum is shown in the left panel of Fig.~\ref{gapspreadn8}). The energy splitting between the two lowest energy states is called the spread $\delta$. In the case of FQHE, the spread should be equal to zero due to the center of mass degeneracy. A necessary condition to be able to distinguish the two lowest energy states is $\delta$ to be smaller than the gap $\Delta$ (defined as the energy difference between the third and the second lowest energy levels). Another necessary condition to claim a Laughlin-like is hosted in this system is related to the quantum number of the two lowest energy states: If they are associated to a Laughlin-like state, they should be given by the counting principle described in Ref.~\cite{Bernevig-2012PhysRevB.85.075128}.

\begin{figure}[htb]
\includegraphics[width = 0.21\textwidth]{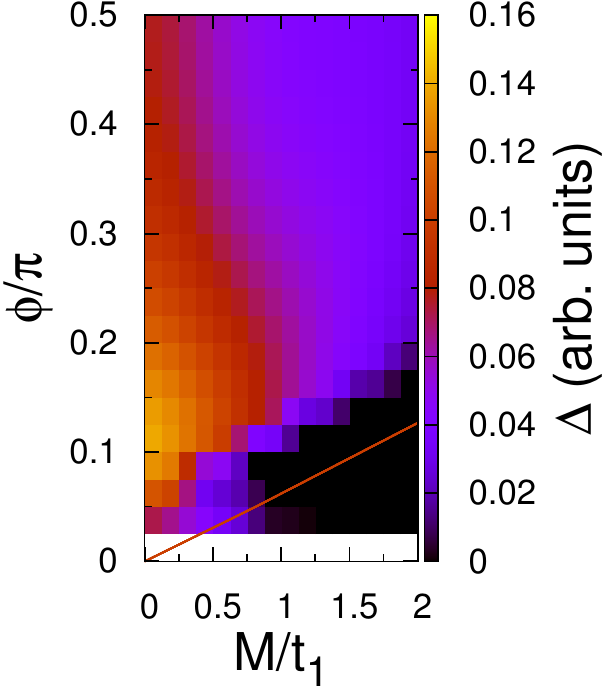}\hspace{0.03\textwidth}
\includegraphics[width = 0.2\textwidth]{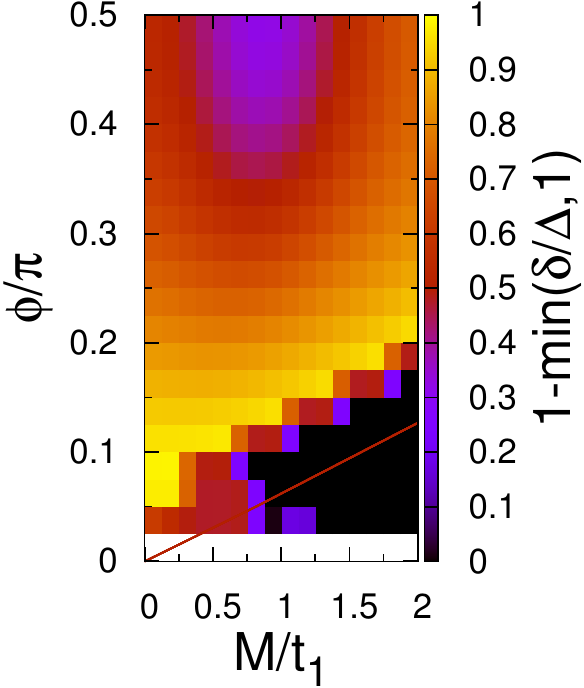}
\caption{{\it Left panel:} The energy gap $\Delta$ as a function of the two tight-binding model parameters $\phi$ and $M/t_1$, for $N=10$ bosons on a $N_x=5, N_y=4$. The gap is set to zero when the two lowest energy states are not in the expected momentum sectors of the Laughlin state, here at $(K_x=0,K_y=0)$ and $(K_x=0,K_y=2)$. The red line denotes the separation between the Chern insulator phase (upper part) and the trivial phase (lower part). {\it Right panel:} The corresponding spread $\delta$ (displayed as $1-{\rm min}(\delta/\Delta,1)$) as a function of $\phi$ and $M/t_1$. A color value of 1 would correspond to a perfectly twofold degenerate ground state (i.e. $\delta=0$).}
\label{gapspreadn10}
\end{figure}

We have computed the phase diagram when tuning $\phi$ and $M$ at filling $\nu=1/2$ for two different system sizes: $N=8$ on a $N_x=N_y=4$ lattice (Fig.~\ref{gapspreadn8}) and $N=10$ on a $N_x=5, N_y=4$ lattice (Fig.~\ref{gapspreadn10}). We show both the gap $\Delta$ and the spread $\delta$ (actually $1-{\rm min}(\delta, \Delta, 1)$ such that we have $1$ when the spread is $0$ and $0$ if $\delta > \Delta$). When the band structure parameters are set to values leading to a trivial band, we clearly see that the FCI phase completely disappears. In the non-trivial region, both system sizes suggest a robust Laughlin-like state around $M=0$ and $\phi=0.11-0.12\pi$.

\begin{widetext}
\begin{center}
\begin{figure}[htb]
\includegraphics[width = 0.27\textwidth]{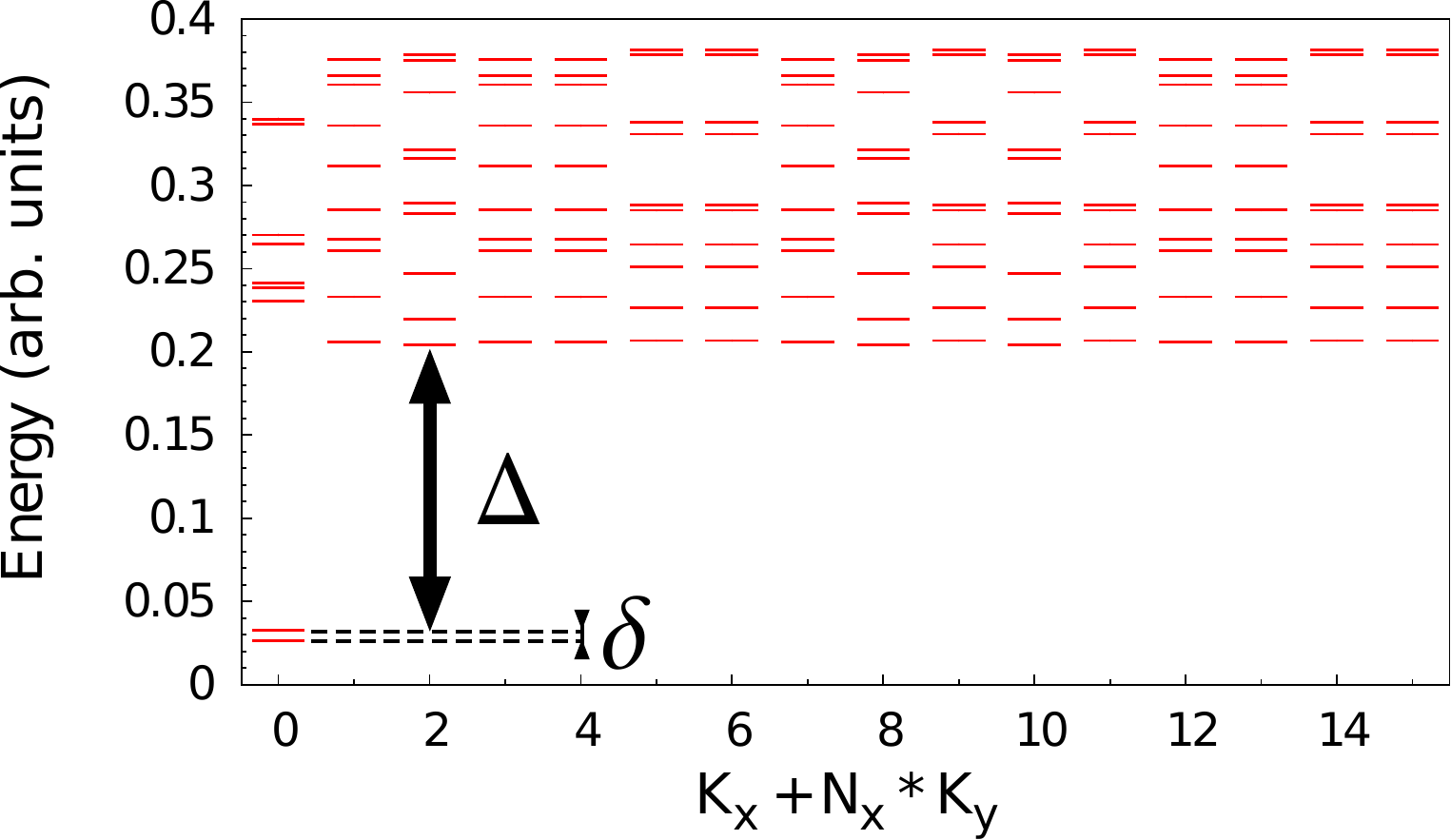}\hspace{0.05\textwidth}
\includegraphics[width = 0.30\textwidth]{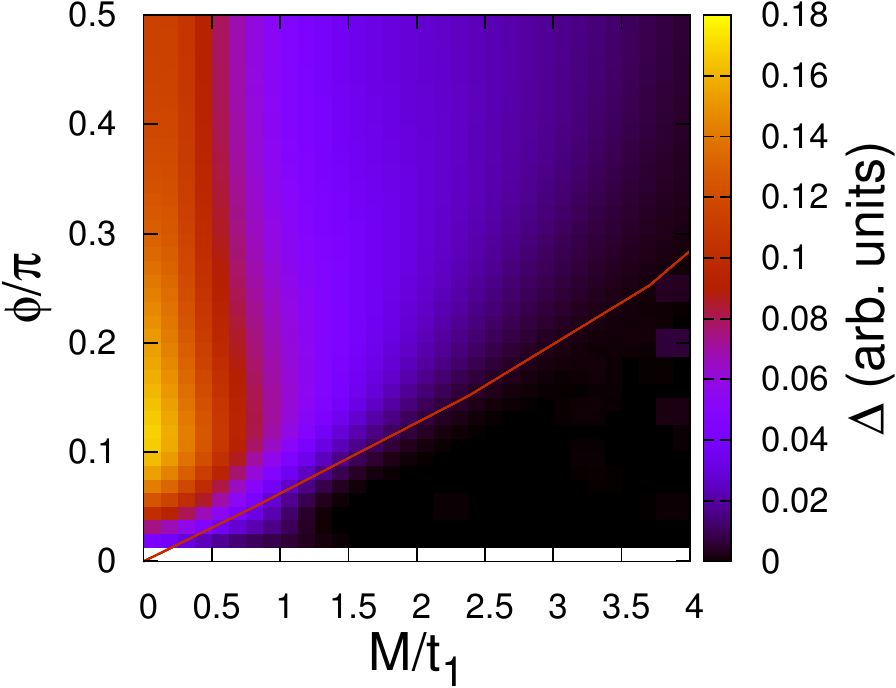}\hspace{0.05\textwidth}
\includegraphics[width = 0.29\textwidth]{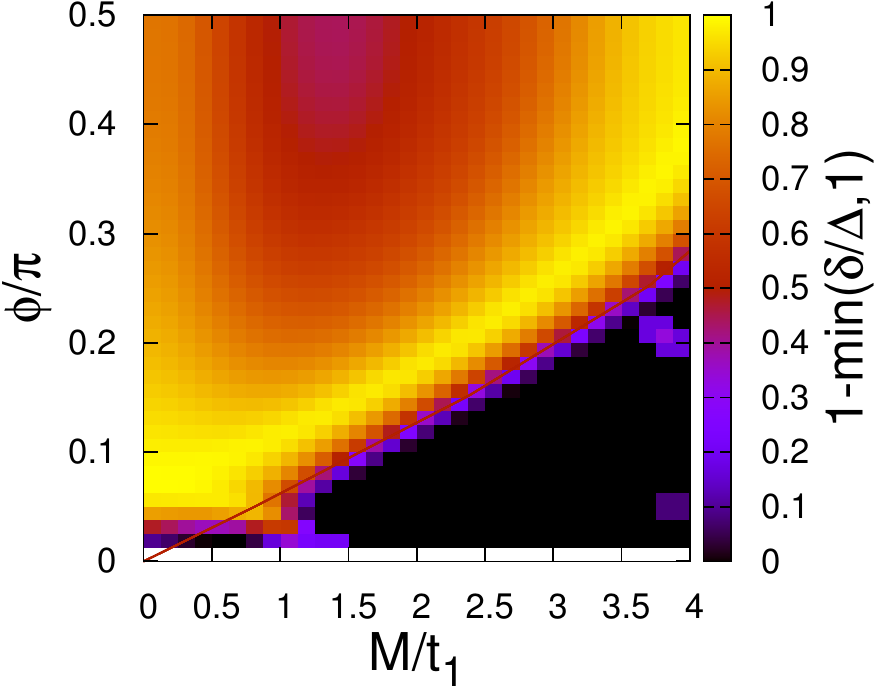}
\caption{{\it Left panel:} Typical low energy spectrum for the FCI Haldane model as a function of the linearized two-dimensional momentum $(k_x,k_y)$. Here we have set $N=8$, $N_x=N_y=4$, $M=0$ and $\phi=0.12 \pi$. The spread $\delta$ is the splitting between the two lowest energy states, corresponding to the twofold degenerate Laughlin states. The gap $\Delta$ is defined as the energy difference between the third lowest energy and the second lowest energy, irrespective of the momentum sector. {\it Middle panel:} The energy gap $\Delta$ as a function of the two tight-binding model parameters $\phi$ and $M/t_1$, for $N=8$ bosons on a $N_x=4, N_y=4$. The gap is set to zero when the two lowest energy states are not in the expected momentum sectors of the Laughlin state, here at $(K_x=0,K_y=0)$ for both states. The red line denotes the separation between the Chern insulator phase (upper part) and the trivial phase (lower part). {\it Right panel:} The corresponding spread $\delta$ (displayed as $1-{\rm min}(\delta/\Delta,1)$) as a function of $\phi$ and $M/t_1$. A color value of 1 would correspond to a perfectly twofold degenerate ground state (i.e. $\delta=0$).}
\label{gapspreadn8}
\end{figure}
\end{center}
\end{widetext}

\subsection{The single-particle background}
We now consider the one-body hamiltonian properties. In the left panel of Fig.~\ref{HaldaneGandB}, we show the non-universal nature of $g$ - the averaged over BZ diagonal element of the quantum distance (FS) metric, more precisely $g A_{\text{BZ}}$ (where $A_{\text{BZ}}$ is the BZ area), is illustrated for  the (two band) Haldane model with fixed parameters $ t_1 = t_2 = 1$. The standard deviations divided by the averaged values $g$ and the standard deviations of the Berry curvatures of the Haldane model are shown in the middle and right panel of Fig.~\ref{HaldaneGandB}. Note that both the relative deviations of the FS metric and those of the Berry curvature are minimal around the point $(M, \phi) = (0, 0.11\pi)$. This is around the same region that we have observed the strongest Laughlin-like state in our finite size numerical calculations.

\begin{figure}[hbt]\centering
\includegraphics[width=0.35\textwidth]{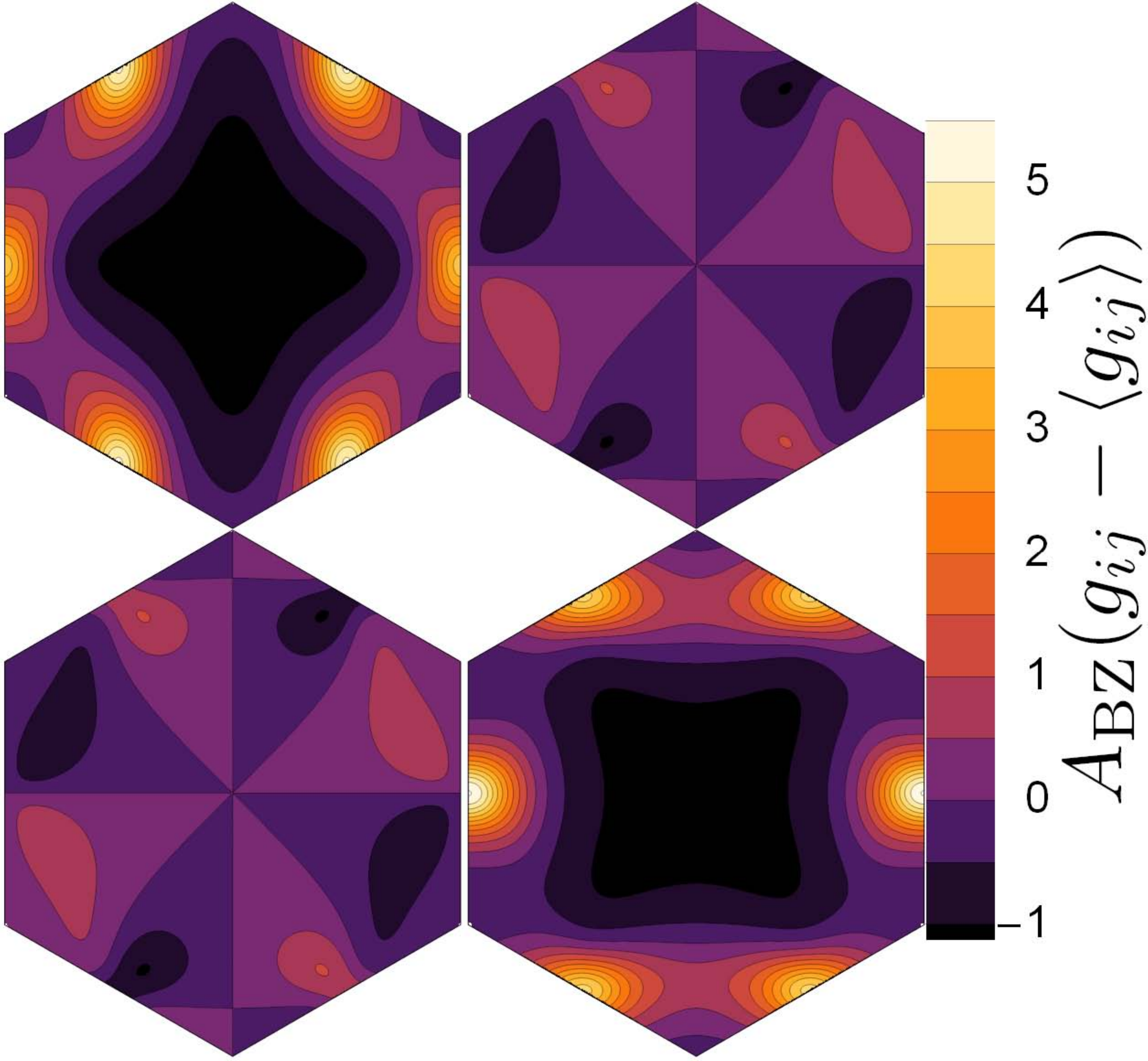}
\caption{\label{particularG} The values of the FS metric with respect to its mean values at $(M,\phi)=(0,0.11\pi)$. The mean values are $A_{\text{BZ}} \langle g_{11} \rangle = A_{\text{BZ}} \langle g_{22} \rangle = 1.13$ and $\langle g_{12} \rangle = \langle g_{21} \rangle = 0$ in the units of $A_{\text{BZ}}|\overline{B}| / 2 = \pi$. In the Figure the graphs are ordered as the metric matrix elements. All four metric matrix elements are zero at $\mathbf{k} = (0,0)$ momentum.}
\end{figure}

\begin{figure}[hbt]\centering
\includegraphics[width=0.27\textwidth]{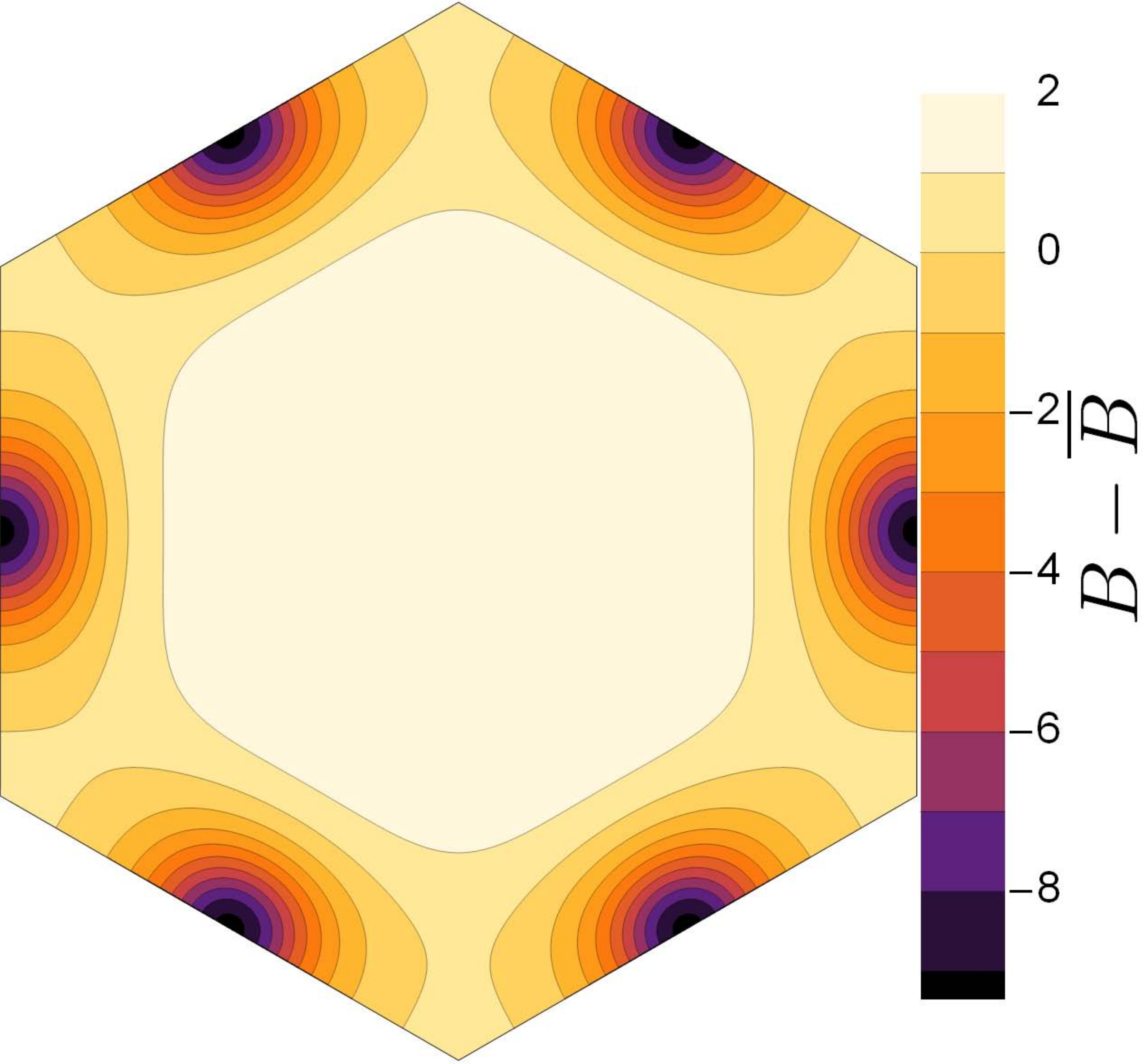}\hspace{0.02\textwidth}
\includegraphics[width=0.17\textwidth]{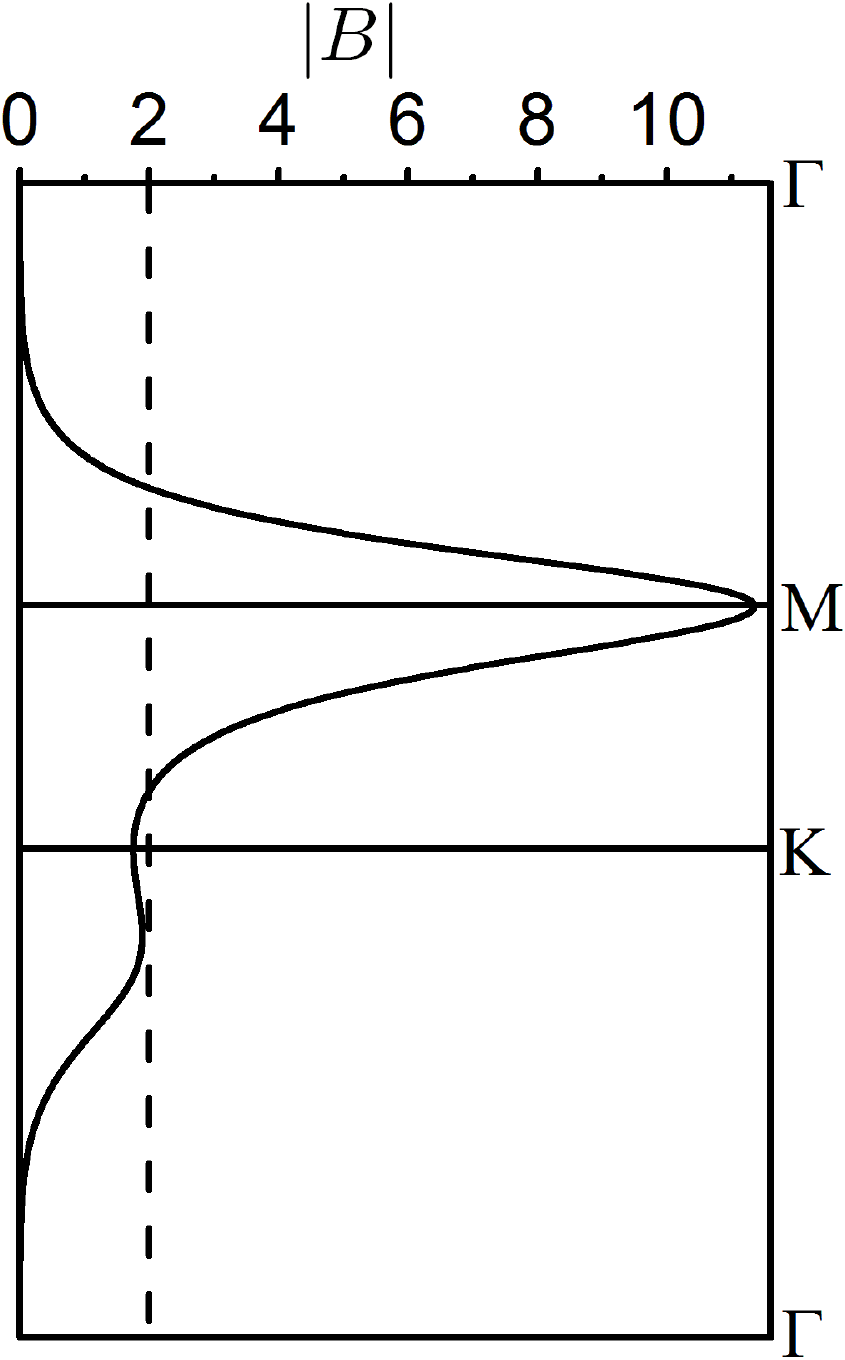}
\caption{\label{particularB} The values of  the Berry curvature in the BZ at $(M,\phi)=(0,0.11\pi)$ with respect to its mean value, $\overline{B} = -2 \pi / A_{\text{BZ}}$, denoted by dash line. Plotted values of B  are in the units of  $\pi / A_{\text{BZ}}$.}
\end{figure}

To provide a more detailed insight of this point $(M, \phi) = (0, 0.11\pi)$, we provide in Fig.~\ref{particularG} the values of the $g_{ij}^{FS}$ tensor (in an orthogonal coordinate system): $g_{11}^{FS}, g_{12}^{FS}, g_{21}^{FS},$ and $g_{22}^{FS}$ as functions  of the Bloch momentum in the Brillouin zone. We also give in Fig.~\ref{particularB} the values of the Berry curvature at the same point in the phase space.

\begin{widetext}
\begin{center}
\begin{figure}[hbt]\centering
\includegraphics[width=0.30\textwidth]{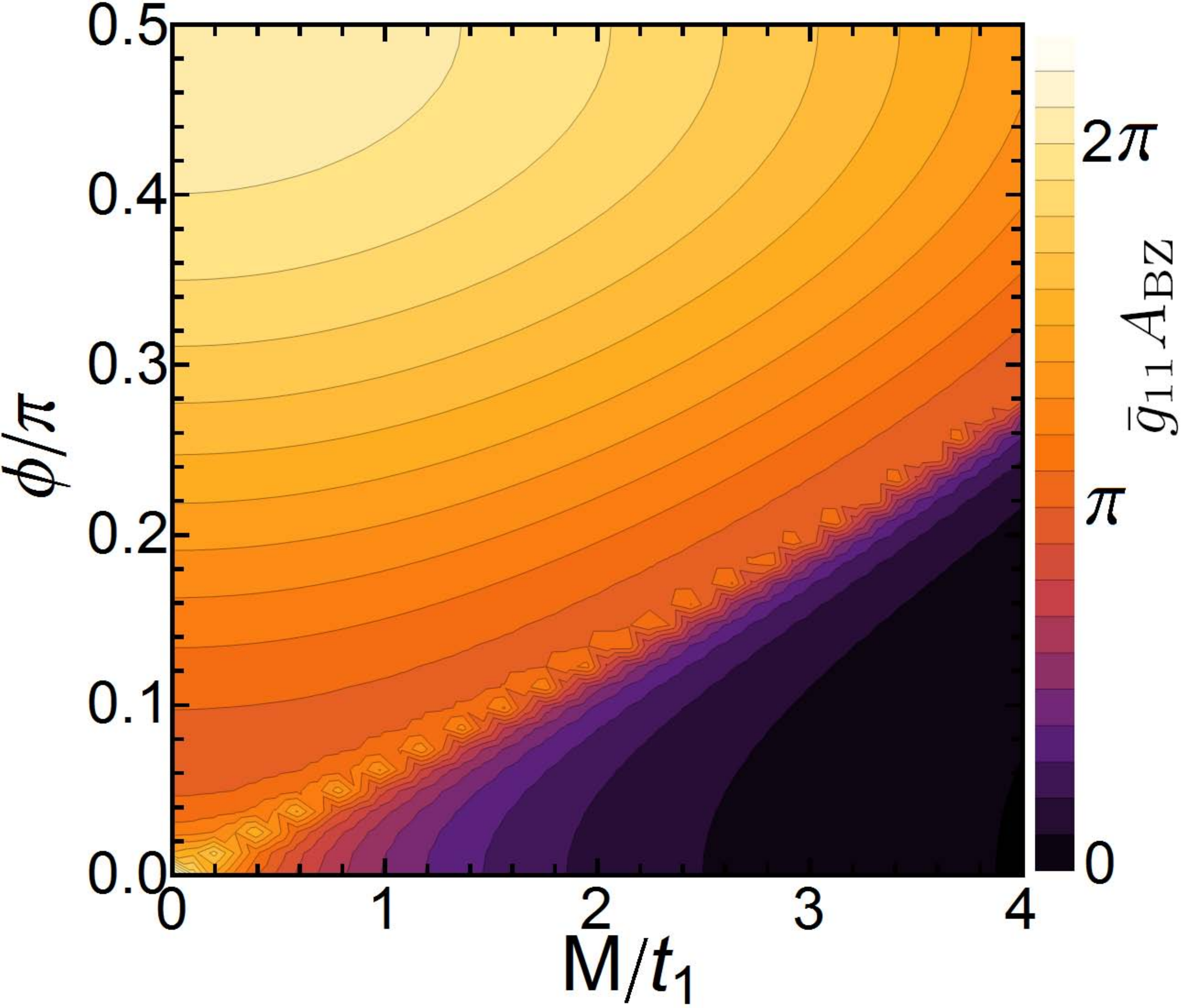}
\hspace{0.02\textwidth}
\includegraphics[width=0.30\textwidth]{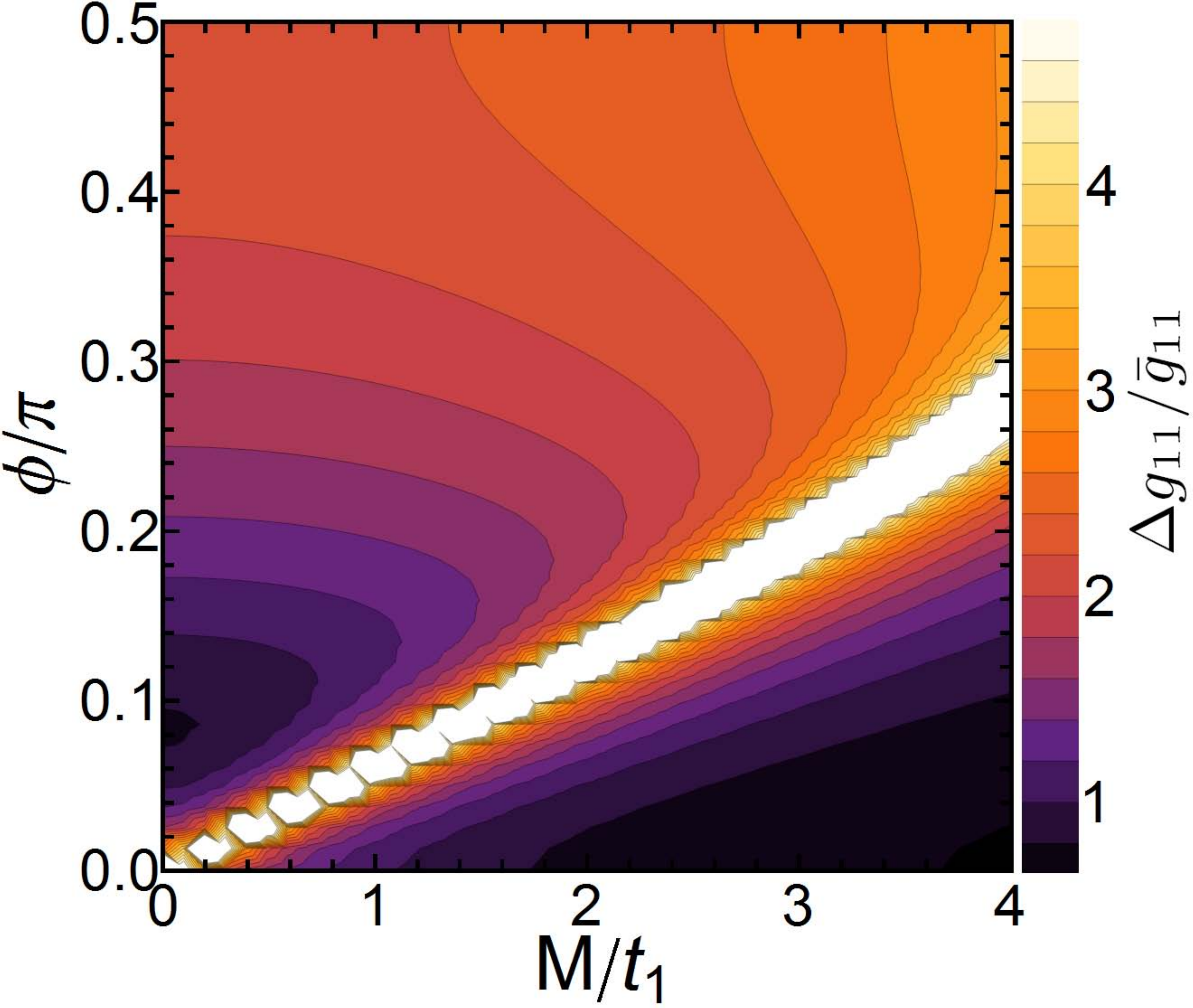}
\hspace{0.02\textwidth}
\includegraphics[width=0.30\textwidth]{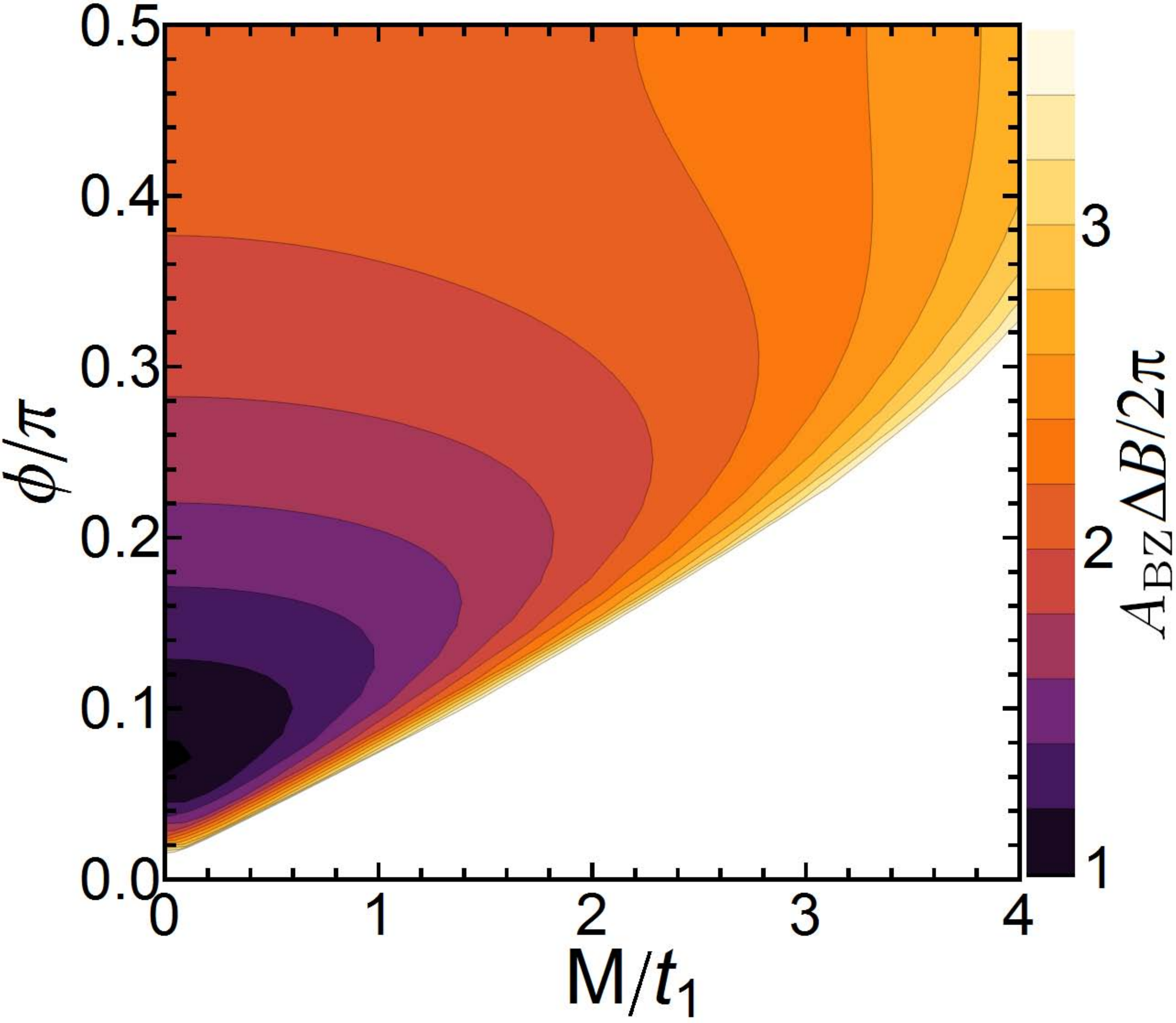}
\caption{\label{HaldaneGandB} {\it Left panel:} $\bar{g}_{11} A_{\text{BZ}}=g A_{\text{BZ}}$ for the Haldane model as a function of $\phi$ and $M$ with $t_1=t_2=1$. {\it Middle panel:} The corresponding deviation of $g_{11}$, i.e. $\Delta g_{11}/\bar{g}_{11}$. {\it Right panel:}  The relative deviation of the  Berry curvature $B$  with respect to its averaged value.}
\end{figure}
\end{center}
\end{widetext}

\begin{figure}[hbt]\centering
\includegraphics[width=0.5\textwidth]{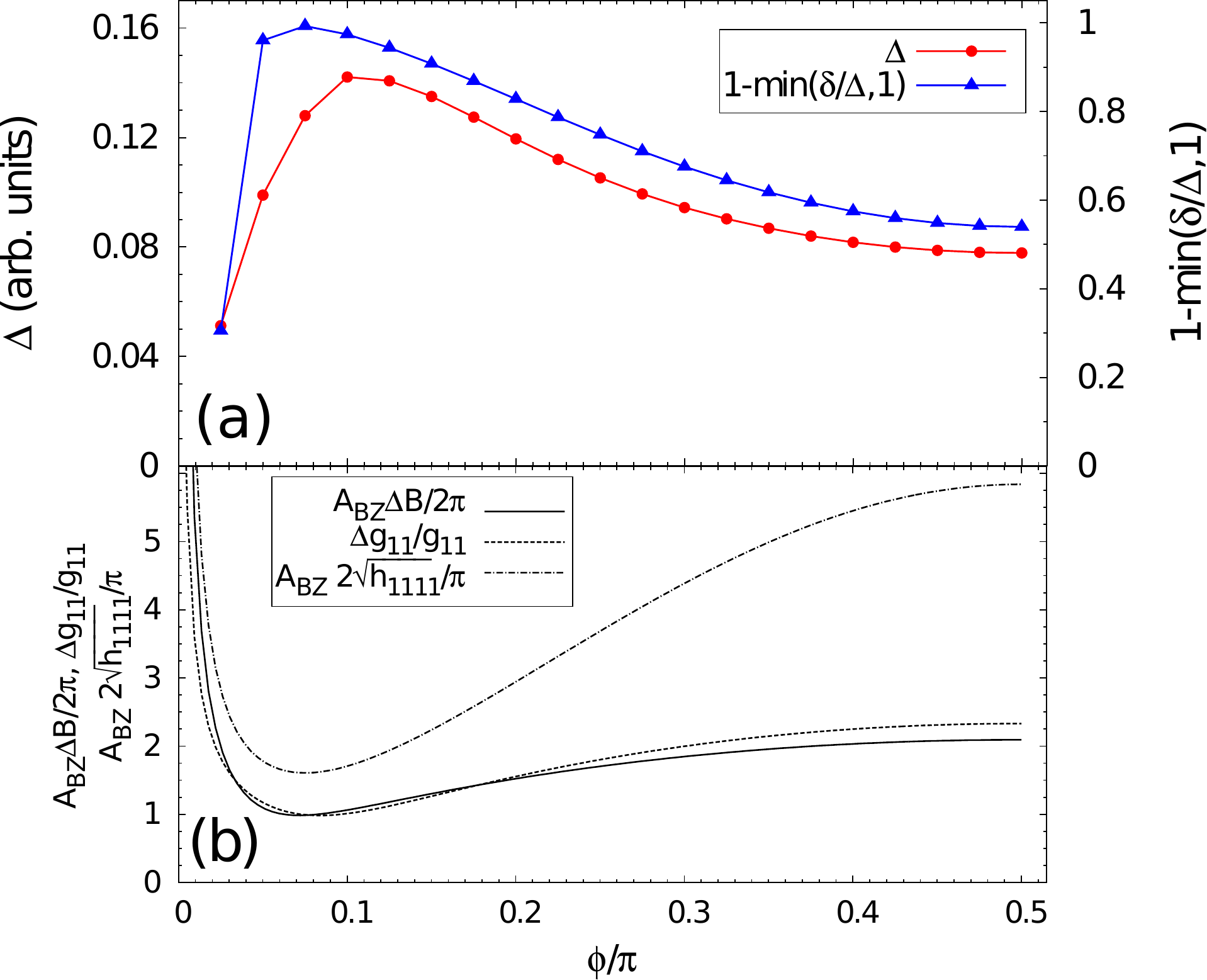}
\caption{\label{FM0} More detailed description of the M = 0 line in the phase space: (a) The values of the gap and spread of the system with 10 particles; (b) Relative deviations of Berry curvature, metric element $g_{11}$ and the coefficient $h\equiv h_{1111}$ defined by Eq. (31).}
\end{figure}

\subsection{Discussion}

When we compare the results of exact diagonalizations  (Fig.~\ref{gapspreadn8} for $N=8$ and Fig.~\ref{gapspreadn10} for $N =10$ particles) with background properties (Fig.~\ref{HaldaneGandB}) we notice that the FCI state is the most probable, with significant gap and two-fold degeneracy of the ground state, whenever both, the variations of Berry curvature and metric are small, and the averaged value of the metric is close to the unimodular requirement i.e. the value of the diagonal element, $\bar{g}_{11} A_{BZ}$, is close to $\pi$ in the left panel of Fig.~\ref{HaldaneGandB}. Nevertheless the results indicate that the FCI state extends and persists for a while beyond this optimal - FQHE region. This occurs despite the understanding that the unimodular requirement is a strong condition for FQHE, and together with Berry curvature variation should influence decay.

In the narrow region along the $M=0$ line of the FCI phase we observe the decline of the gap (Fig.~\ref{FM0}, upper panel) along the increase of the metric average value and variations of Berry curvature and metric (Fig.~\ref{FM0}, lower panel). In this region  we do not expect a formation of charge density wave but possibly a transition into another liquid state - superfluid, the relevance of the magneto-phonon physics and a correlation with long-distance SSF behavior. The liquid transition should be accompanied with the collapse of a magneto-phonon gap, and non-analytical behavior in the SSF. But what we find is persistence of the gap beyond FQHE region (Fig.~\ref{FM0}, upper panel) that is in a qualitative agreement with our analytical estimates of the coefficient of quartic term (Fig.~\ref{FM0}, lower panel and Eqs.~\ref{Ehvalue},~\ref{Eh},~\ref{hFCI} and~\ref{final_quartic}); the coefficient is always positive and grows with the increase of the metric averaged value and variations in the background properties. According to the SMA, the gap is inversely proportional to the coefficient in the long-distance region and, at least, for the magneto-phonon gap we expect its decay but not closing $-$ the behavior that we see in our exact diagonalization results.

If we had approximated the last term in Eq.~\ref{final_quartic} with $n|q|^4 g^2/2$ we would not have such an agreement with exact diagonalization results; namely this approximation predicts phase transition before reaching $\phi = \pi/2$ at $2 g/\overline{B}  = m $. Therefore the persistence of the FCI signatures in numerical results is likely a consequence of variations in metric and other background properties.

Thus we find evidence that a FCI state may persist and sustain a large deviation from the (unimodular) metric of the FQHE state. An example of the metric deviation  in the conventional FQHE we find in the experiments with tilted field~\cite{Xia-NaturePhys.7.845}. Though the analogy is not complete, because of the anisotropy in the FQHE case, there the determinant of the (external, one-particle) metric deviates from value one due to an effective increase of the probability to find electrons in the plane normal to the external magnetic field without tilt~\cite{Papic-PhysRevB.87.245315}. Thus as in the FQHE case~\cite{Mulligan-PhysRevB.82.085102}, a new physics and further broadening of the definition of the FQHE phenomena may occur in the FCI physics. Unfortunately in the FCI case, at present, we are limited by system sizes in the exact diagonalizations to further explore this phenomenon. That the departure from the FQHE case is non-trivial and important is also underlined by the observation that in the phase diagram of the model we considered having the metric near the FQHE value seems sufficient for the presence of  the two-fold degeneracy of the ground state although this does not guarantee a good size of the gap. This can be seen by comparing the averaged values of the metric (Fig.~\ref{HaldaneGandB}, left panel) with the spread characterization of the phase diagram (Figs.~\ref{gapspreadn8} and~\ref{gapspreadn10}, right panel) .

\section{Conclusions}\label{Section:Conclusions}
Based on the SSF calculations we studied the role of band geometry in the context of the FCI physics. We identified that the averaged over BZ FS metric plays the role of the quantum distance metric in the long-wave length domain based on the calculations of the single particle part of the projected to the band SSF. We discussed the behavior of the complete projected SSF in a mean-field framework, and whether and under which conditions the scenario of Ref.~\cite{girvin-PhysRevB.33.2481} is possible in the FCI context. We illustrated the role of the band geometry in the phase diagram of interacting bosons that live on the lattice of the Haldane model. The Laughlin $\nu=1/2$ bosonic FCI state is the most pronounced for the FQHE value of the metric (unimodular requirement) when variations over the BZ of the FS metric and Berry curvature are minimal. But the Laughlin-like phase persists even with the metric increase  in an agreement with the mean field treatment of the long-distance physics of FCI. Further investigations are necessary which may provide us also with reasons for occurrence and stability of FCI states.

\section{Acknowledgment}
We would like to thank A. Bernevig, M. Goerbig, F.D.M. Haldane, Z. Papi\'c, and R. Roy for discussions. This work was supported by the Serbian Ministry of Education and Science under projects No. ON171017 and ON171027. The authors also acknowledge support form the bilateral MES-CNRS 2011/12 program. N.R. was supported by NSF CAREER DMR-095242, ONR-N00014-11-1-0635, ARMY-245-6778, MURI-130-6082, Packard Foundation, and Keck grant. M.V.M. was supported by ONR-N00014-11-1-0635.

\appendix
\section{FCI state-single particle contribution}\label{Appendix:SSPContribution}
Here we analyze the expansion in small momentum  to the quartic order of the expression in Eq.~\ref{no_expansion_single_part} that represents the single particle contribution to the projected SSF. To fourth order with  assumed summations on repeated indices $i,j,k$
\begin{eqnarray}
&&u_{\alpha,p-q}= u_{\alpha,p} - q_i \partial_i u_{\alpha,p} + \frac{q_i q_j}{2} \partial_i \partial_j u_{\alpha,p} - \nonumber \\
&& \frac{q_i q_j q_k}{3!} \partial_i \partial_j \partial_k u_{\alpha,p} + o(q^4),
\end{eqnarray}
and therefore
\begin{eqnarray}
&&u_{\alpha_1,p-q}u_{\alpha_2,p-q}^{*}= u_{\alpha_1,p}u_{\alpha_2,p}^{*} - q_i u_{\alpha_1,p}\partial_i u_{\alpha_2,p}^{*} + \nonumber \\
&& \frac{q_i q_j}{2}u_{\alpha_1,p} \partial_i \partial_j u_{\alpha_2,p}^{*}-
\frac{q_i q_j q_k}{3!} u_{\alpha_1,p} \partial_i \partial_j \partial_k u_{\alpha_2,p}^{*} \nonumber \\
&&-q_i \partial_i u_{\alpha_1,p}\; u_{\alpha_2,p}^{*} + q_i q_j \partial_i u_{\alpha_1,p} \partial_j u_{\alpha_2,p}^{*} \nonumber \\
&&- q_i \frac{q_k q_j}{2}
\partial_i u_{\alpha_1,p}\partial_i \partial_j u_{\alpha_2,p}^{*} \nonumber \\
&&+ \frac{q_i q_j}{2} \partial_i \partial_j u_{\alpha_1,p} \; u_{\alpha_2,p}^{*} -
\frac{q_i q_j}{2} q_k \partial_i \partial_j u_{\alpha_1,p} \; \partial_k u_{\alpha_2,p}^{*} \nonumber \\
&&- \frac{q_i q_j q_k}{3!} \partial_i \partial_j \partial_k u_{\alpha_1,p} \; u_{\alpha_2,p}^{*} + o(q^4).
\end{eqnarray}
Because $u_{\alpha,p}u_{\alpha,p}^{*} = 1$ and therefore $u_{\alpha,p}\partial_i u_{\alpha,p}^{*} + \partial_i u_{\alpha,p} \; u_{\alpha,p}^{*} = 0$ we have to second order
\begin{equation}
\langle \tilde{\rho}_{-q} \tilde{\rho}_{q}\rangle|_{\text{single}}= n + o(q^2).
\end{equation}
To the second order we have
\begin{eqnarray}
&& u_{\alpha_1,p}^{*} u_{\alpha_1,p-q} u_{\alpha_2,p-q}^{*} u_{\alpha_2,p} \approx 1 +
\nonumber \\
&& \frac{q_i q_j}{2} [ u_{\alpha,p} \partial_i \partial_j u_{\alpha,p}^{*} +
\partial_i \partial_j u_{\alpha,p} \; u_{\alpha,p}^{*} \nonumber \\
&& + 2 \; \partial_i u_{\alpha_1,p} \; u_{\alpha_1,p}^{*} \; u_{\alpha_2,p} \partial_i u_{\alpha_2,p}^{*}].
\end{eqnarray}
Due to
\begin{eqnarray}
&& A_{ij} [ u_{\alpha,p} \partial_i \partial_j u_{\alpha,p}^{*} +
\partial_i \partial_j u_{\alpha,p}\; u_{\alpha,p}^{*} \nonumber \\
&& + 2 \partial_i u_{\alpha_1,p} \partial_j u_{\alpha_1,p}^{*}] = 0,
\end{eqnarray}
for any symmetric $A_{ij}$ we have
\begin{equation}
\langle \tilde{\rho}_{-q} \tilde{\rho}_{q}\rangle |_{\text{single}} = n - q_i q_j \sum_p g_{ij}^{FS}(p) n_p + o(q^3),
\end{equation}
where
\begin{eqnarray}
&& g_{ij}^{FS}(p) = \frac{1}{2}  [ \partial_i u_{\alpha_1,p} \partial_j u_{\alpha_1,p}^{*} + \partial_j u_{\alpha_1,p} \partial_i u_{\alpha_1,p}^{*} -
 \nonumber \\ && \partial_i u_{\alpha_1,p} u_{\alpha_1,p}^{*} u_{\alpha_2,p}\partial_j u_{\alpha_2,p}^{*} - \partial_j u_{\alpha_1,p} u_{\alpha_1,p}^{*} u_{\alpha_2,p}\partial_i u_{\alpha_2,p}^{*}] \nonumber \\
\end{eqnarray}
is the Fubini-Study metric. To the third order we find
\begin{eqnarray}
&& u_{\alpha_1,p}^{*} u_{\alpha_1,p-q} u_{\alpha_2,p-q}^{*} u_{\alpha_2,p} \approx 1 + q_i q_ j g_{ij}^{FS}(p)
\nonumber \\
&& - \frac{q_i q_j q_k}{3!} [ u_{\alpha,p} \partial_k \partial_i \partial_j u_{\alpha,p}^{*} +
\partial_k \partial_i \partial_j u_{\alpha,p}\; u_{\alpha,p}^{*} \nonumber \\
&& + 3 \partial_i u_{\alpha_1,p} u_{\alpha_1,p}^{*} u_{\alpha_2,p}\partial_k \partial_j u_{\alpha_2,p}^{*}+ \nonumber \\
&& 3 \partial_i \partial_k u_{\alpha_1,p} u_{\alpha_1,p}^{*} u_{\alpha_2,p} \partial_j u_{\alpha_2,p}^{*}] + o(q^4).
\end{eqnarray}
Again differentiating $u_{\alpha,p}u_{\alpha,p}^{*} = 1$ three times we have
\begin{eqnarray}
&& A_{ijk} [ u_{\alpha,p} \partial_i \partial_j \partial_k u_{\alpha,p}^{*} +
\partial_i \partial_j \partial_k u_{\alpha,p} \; u_{\alpha,p}^{*} \nonumber \\
&& + 3 \partial_i \partial_k u_{\alpha_1,p} \partial_j u_{\alpha_1,p}^{*} \nonumber \\
&& + 3 \partial_i  u_{\alpha_1,p} \partial_j\partial_k u_{\alpha_1,p}^{*} ] = 0,
\end{eqnarray}
and therefore
\begin{eqnarray}
&& u_{\alpha_1,p}^{*} u_{\alpha_1,p-q} u_{\alpha_2,p-q}^{*} u_{\alpha_2,p} \approx 1 + q_i q_ j g_{ij}^{FS}(p)+
\nonumber \\
&& \frac{q_i q_j q_k}{2} [ \partial_k  u_{\alpha,p} \partial_i \partial_j u_{\alpha,p}^{*} +
 \partial_i \partial_j u_{\alpha,p} \partial_k u_{\alpha,p}^{*} \nonumber \\
&& -  \partial_i u_{\alpha_1,p} u_{\alpha_1,p}^{*} u_{\alpha_2,p}\partial_k \partial_j u_{\alpha_2,p}^{*}+ \nonumber \\
&& - \partial_i \partial_k u_{\alpha_1,p} u_{\alpha_1,p}^{*} u_{\alpha_2,p} \partial_j u_{\alpha_2,p}^{*}] = \nonumber \\ &&  1 + q_i q_ j g_{ij}^{FS}(p) + \frac{q_i q_j q_k}{2} \partial_k g_{ij}^{FS}(p).
\end{eqnarray}

\section{SSF definition}\label{Appendix:SSFDefinition}
By expanding the expression in Eq.~\ref{no_def} we immediately see that the quadratic contribution in $q$ will be
\begin{equation}
\sum_{i,j} |q|^2 (R_i^{\dagger} R_j + R_i  R_j^{\dagger}).
\end{equation}
  Therefore even when the condition in Eq.~\ref{lcondition} is applied we have a non-zero contribution to the quadratic order. The expression in Eq.~\ref{no_def} does not correspond to the usual definition of the SSF as in Ref.~\onlinecite{girvin-PhysRevB.33.2481}, and only after an additional subtraction it reproduces the well-known behavior in the classical (continuum) FQHE.

We will illustrate and explain the source of this discrepancy in the coordinate representation of the continuum FQHE. First, by using the Eqs.~\ref{prodensity}-\ref{singleSSF} we see that~\cite{girvin-PhysRevB.33.2481}
\begin{equation}
\tilde{s}(q) = s(q) - n (1 - \exp\{- \frac{|q|^2}{2}\}). \label{srelation}
\end{equation}
(This is most easily seen in the first quantization picture of the many-body problem considering the action of the translation operator in the LLL, $\exp\{\mathrm{i} q \partial/\partial z\}$; we rederived  Eq.~\ref{srelation} in Section~\ref{Subsection:SSF}  in the second quantization as a step towards the discussion of the FCI state.)

To extract the quadratic contribution to $\tilde{s}(q)$ we consider the expansion of the unprojected SSF, $s(q)$, and the following correlator of unprojected densities,
\begin{equation}
\frac{1}{V} \sum_{i,j} \langle \exp\{ \mathrm{i} \mathbf{q} \cdot \mathbf{r}_i\} \exp\{ - \mathrm{i} \mathbf{q} \cdot \mathbf{r}_j\} \rangle.
\end{equation}
If we assume the conservation of the angular momentum and use the complex notation of the LLL for the quadratic term, after the subtraction of ``self-terms" i.e. these generated by $ \sum_{i} \langle \exp\{ \mathrm{i} \mathbf{q} \cdot \mathbf{r}_i\} \rangle $, we have for the quadratic term the following  expressions,
\begin{eqnarray}
&& \frac{1}{V} \sum_{i,j} \langle  \mathbf{q} \cdot \mathbf{r}_i  \mathbf{q} \cdot \mathbf{r}_j \rangle = \nonumber \\
&& \frac{1}{V} \frac{|q|^2}{4} \sum_{i,j} \langle  z_i^{*} z_j + z_j^{*} z_i \rangle = \nonumber \\
&& \frac{1}{V} \frac{|q|^2}{2} \sum_{i,j} \langle  \frac{\partial}{\partial z_i} z_j + \frac{\partial}{\partial z_j} z_i \rangle = \nonumber \\
&& n |q|^2 \label{string}.
\end{eqnarray}
To get the final expression we used the properties of the LLL functions (which are holomorphic up to the Gaussian factor) and the property of the homogeneity of the ground state i.e. that its holomorphic part is annihilated by the $ \sum_{i}  \partial/\partial z_i$  operator (Eq.~\ref{lcondition}) in the momentum representation).

The substitution of the expression~(\ref{string}) for $s(q)$ in Eq.~\ref{srelation} would lead to a non-zero, quadratic in $q$ contribution to $\tilde{s}(q)$. This difference in the value that we have  for $s(q)$ must stem from a difference in the subtractions: the one used in Ref.~\onlinecite{girvin-PhysRevB.33.2481} and the other, when in a static correlator the ground state values of the two densities are subtracted (``self-terms"), implied by the expression of Eq.~\ref{no_def}. Namely, in the classical reference the subtraction (i.e. a procedure to avoid divergences) is introduced for $s(q)$ at any $q \neq 0$ as
\begin{equation}
s(q) = n + n^2 \int \mathrm{d} \mathbf{r}  (g(r) - 1) \exp\{-\mathrm{i}\mathbf{q}\cdot\mathbf{r}\} \label{sub_gmp},
\end{equation}
because the combination, $(g(r) - 1)$, leads to the absence of the divergences for large $r$.  For example, in the integer QH case we have (as an exact expression) $ g(r) = 1 - \exp\{- r^2/2\}$, and this leads to the usual, well-known behavior $ s(q) \approx n |q|^2/2$ and $\tilde{s}(q) = 0$. In Ref.~\onlinecite{girvin-PhysRevB.33.2481} it was shown that $ s(q) \approx n |q|^2/2$ for any liquid ground state of the system that conserves the angular momentum and particle number. The difference between this conclusion and the result in Eq.~\ref{string} stems from different subtraction procedures, and can be traced back to two different definitions of the SSF. The first definition is given  in Eq.~\ref{sub_gmp} and defines a static limit of the time ordered density-density correlator, and the second one describes a static correlator from which ``self-terms" are subtracted (as in the expression of Eq.~\ref{no_def} in the momentum representation in the projected case).

\bibliography{fcissf}

\end{document}